\documentclass[onecolumn]{aastex6}

\newcommand{\kms}{${\rm km~s}^{-1}$}

\slugcomment{Accepted for publication by ApJ}
\shorttitle{RR Lyrae stars in the VOD}
\shortauthors{A. K. Vivas et al.}

\begin{document}

\title{Disentangling the Virgo Overdensity with RR Lyrae stars}

\author{A. Katherina Vivas\altaffilmark{1}}
\affil{Cerro Tololo Interamerican Observatory, Casilla 603, La Serena, Chile}

\author{Robert Zinn}
\affil{Department of Astronomy, Yale University, PO Box 208101, New Haven, CT 06520-8101, USA}

\author{John Farmer}
\affil{Department of Physics, University of Chicago, 5720 South Ellis Avenue, Chicago IL 60637, USA}

\author{Sonia Duffau}
\affil{Millennium Institute of Astrophysics, Chile \\
Instituto de Astrof{\'\i}sica, Facultad de F{\'\i}sica, Pontificia Universidad Cat\'olica de Chile, Av. Vicu\~na Mackenna 4860, 782-0436 Macul, Santiago, Chile}

\and

\author{Yiding Ping}
\affil{Department of Astronomy, Yale University, PO Box 208101, New Haven, CT 06520-8101, USA}

\altaffiltext{1}{kvivas@ctio.noao.edu}

\begin{abstract}
We use a combination of spatial distribution and radial velocity to search for halo sub-structures in a sample of 412 RR Lyrae stars (RRLS) that covers a $\sim 525$ square degrees region of the Virgo Overdensity (VOD) and spans distances from the Sun from 4 to 75 kpc.  With a friends-of-friends algorithm we identified six
high significance groups of RRLS in phase space, which we associate mainly with the VOD and with the Sagittarius stream. Four other groups
were also flagged as less significant overdensities.
Three high significance and 3 lower significance groups have distances between $\sim 10$ and 20 kpc, which places them with the distance range attributed by others to the VOD.  The largest of these is the Virgo Stellar Stream (VSS) at 19 kpc, which has 18 RRLS, a factor of 2 increase over the number known previously.  While these VOD groups are distinct according to our selection cirteria, their overlap in position and distance, and, in a few cases, similarity in radial velocity are suggestive that they may not all stem separate accretion events.  Even so, the VOD appears to be caused by more than one overdensity. 
The Sgr stream is a very obvious feature in the background of the VOD at a mean distance of 44 kpc. Two additional high significant groups were detected at distances $>40$ kpc.  Their radial velocities and locations
differ from the expected path of the Sgr debris in this part of the sky, and they are likely to be remnants of other accretion events.  
\end{abstract}


\section{Introduction}

With the advent of large scale surveys in the last two decades, several low-surface brightness, large-scale
structures, extending hundreds
to few thousand of square degrees in the sky, have been discovered in the halo
of the Milky Way \citep[][and references therein]{newberg16}. 
Contrary to the case of the Sagittarius (Sgr) dwarf spheroidal (dSph)
galaxy, whose debris has been found in well defined tidal streams wrapping around the 
full sky \citep{majewski03}, these structures have more of a {\sl cloud} morphology and their progenitors
are still unknown \citep{grillmair16}. They are nonetheless interpreted as the remains of galaxies that 
were disrupted during the merging with the Milky Way \citep[see for example][]{helmi11}.

At low-galactic latitudes, two such cloud-like structures have been found: the Monoceros ring 
\citep{newberg02} and the Triangulum-Andromeda overdensity \citep{rocha04}. 
In both cases the interpretation of an extra-galactic origin for these features is not clear
and their possible relationship with the galactic disk is still under investigation 
\citep[see][]{price15,xu15}.
At high galactic latitudes and far from the disk, there are several other structures: the Virgo overdensity 
\citep[VOD,][]{vivas01,newberg02,duffau06,juric08,bonaca12,duffau14}, the Hercules-Aquila
cloud \citep{belokurov07,watkins09,simion14}, the Pisces overdensity 
\citep{sesar07,watkins09,kollmeier09,sesar10b} and the Eridanus-Phoenix overdensity \citep{li16}.  Although their interpretation as merger event, or events, is less ambiguous, the specific 
merging history is still unknown. The existence of more than one kinematic substructure in some of these clouds 
\citep{sesar10b,newberg07,vivas08,duffau14} hints that they may contain either separate merger events, or 
different wraps of debris material from the same event. Currently available data cannot answer this issue.
In addition, no progenitors have been associated with any of these cloud-like structures. 
One possible explanation is that these clouds are
formed  by debris of satellites in highly eccentric orbits whose stars pile-up near the apocenter of their orbit 
as they spend a longer time near that
location \citep{johnston16}. The true extensions of the clouds
are also uncertain.  A connection between several
cloud-like sub-structures, which are located far from each other, has also
been proposed. \citet{li16} suggest, for example, that the VOD, the Hercules-Aquila cloud and the
Eridanus-Phoenix overdensity, all located at a distance of roughly $\sim20$ kpc from the Sun, may
share a common polar-type orbit around the Milky Way and hence, may be related.
Clearly, more investigation is needed in order to understand the origin of these
structures, their relationship with each other and their role in the formation of galaxies 
like our own Milky Way. Connection between different sub-structures in the halo is not a new idea and
``streams of galaxies'' have been proposed before \citep[e.g. the Fornax stream,][]{lyndenbell95}, although
proving their real connection has not been an easy task \citep{piatek07}.

Detailed studies in the VOD have proven to be very difficult mainly because of its very
large expanse on the sky \citep[up to 3,000 sq. deg.,][]{bonaca12}, 
and extension along the line of sight \citep[$\sim 5-20$ kpc,][]{juric08}. 
These investigations often provide conflicting 
results. For example, using deep photometry, \citet{jerjen13} identified
a main sequence population at 23 kpc near the center of the VOD. 
They did not find clear evidence of a population at 19 kpc, which is where the main overdensity of RR Lyrae stars (RRLS) is located \citep{vivas06,duffau14}. 
Similarly, \citet{brink10} found a peak in the velocity 
distribution of main sequence stars which is not coincident with the main peaks found by \citet{newberg07}
using F-turnoff stars and by \citet{duffau06,duffau14} using RRLS.
Both the works of \citet{brink10} and \citet{jerjen13} 
were based in detailed studies of very small regions (of the order of arcmin) in an overdensity which covers a few thousand of square 
degrees. Thus, rather than conflicting results, each study may be providing pieces 
for what seems to be a complex and large puzzle. 

Here we present an expanded kinematic study in the VOD region
covering $\sim 525$ sq degrees using RRLS as tracers,
building on the earlier investigation by \citet{duffau14} of the area of the sky covered by the QUEST survey, which is $\sim 5$ times smaller
than our present study.
RRLS are abundant in the VOD, and indeed, this type of stars provided the first indication of an overdensity \citep{vivas01}. The main advantage of using
RRLS as tracers of the old population of halo sub-structures is that they are excellent standard candles. RRLS are pulsating variable stars located on the Horizontal Branch in the Hertzsprung-Russell diagram; consequently their luminosities span only a small range, allowing the precise determination of their distances.
Period-Luminosity relationships have been found for RRLS in all bands from optical to infrared bands \citep[e.g.][]{cacciari03,caceres08,neely15}.
Thus, a combination of good 3D positioning of RRLS (coordinates and distance) with radial velocities 
provide a powerful way to disentangle the kinematic groups within the region of the VOD. 
Catalogs of RRLS covering large portions of the sky 
\citep{drake13a,drake13b,sesar13a,zinn14,torrealba15} have been made available in the last 
few years, providing the targets for this type of investigation. 

\section{Data}

\subsection{Catalogs of RR Lyrae Stars}

Our primary source of RRLS comes from La Silla-QUEST survey 
\citep[LSQ,][]{zinn14}, a 840 sq deg survey covering a large portion of the VOD. This survey provides a catalog of 1,372 RRLS with 
well-sampled light curves (mean number of epochs per star is $\sim 70$) and covering a range of
distances between 5 and 80 kpc from the Sun. An overdensity of RR Lyrae stars in the 
Virgo region, between 5 and 20 kpc, is easily distinguished in this survey
\citep[see Figure 14 in][]{zinn14}. 
Here, we concentrate in the region where the
VOD seems to be present according to the density maps shown 
in \citet{zinn14}, specifically $175\degr < \alpha
< 210\degr$ and $-10\degr < \delta < +10\degr$. 
In this region, there are 1,054 RRLS in the catalog.

The LSQ survey has partial overlap in the Virgo region with other large scale surveys of RRLS, including
QUEST \citep{vivas04}, the Catalina Real-Time Transient Survey 
\citep[CRTS,][]{drake13a,drake13b,drake14} and the Lincoln Near-Earth Asteroid Research program
\citep[LINEAR][]{sesar13a}. The QUEST survey covers a relatively narrow band of declination, between $\sim -4\degr$
to $0\degr$; it has a lower number of epochs ($<40$) than LSQ, which is sufficient for
obtaining good parameters of the light curves. 
Our previous study on kinematic groups in the VOD was based on this survey
\citep{duffau14}.
The CRTS catalogs of Drake et al. provide 
$\sim 14,000$ RRLS in all the sky between declination $-20\degr$ and $+60\degr$ with a large number of epochs (a few hundred). The catalog contains RRLS up to 
$\sim 60$ kpc \citep{drake13a}.
Finally, LINEAR covers $\sim 8,000$ sq degrees of the sky. Although this is a shallower survey
(reaching only to 30 kpc), it covers the range of distance of the VOD. 
Since none of those surveys is complete by itself, we cross-matched all catalogs,
and obtained a list of 1,410 unique RRLS in the VOD region. For stars in common between 
different
catalogs, we adopted preferentially the light curve properties measured by LSQ
because it is the deepest of these catalogs. Both the amplitude
of the light curve and the time at maximum light are necessary to calculate the systemic radial 
velocity of the stars, as explained later. The catalogs of \citet{drake13b} and \citet{drake14} do not provide 
the time at maximum light for the catalogued  RRLS. For those stars, we downloaded the light curve data
from the CRTS DR2 webpage\footnote{\url http://nesssi.cacr.caltech.edu/DataRelease/} and fit a template to the light
curve in order of obtaining the HJD$_0$ \citep[see][]{vivas08}. 
 
\floattable
\begin{deluxetable*}{rrccccc}[htb!]
\tablecolumns{7}
\tablewidth{0pc}
\tablecaption{Heliocentric radial velocity for the RR Lyrae stars from individual SDSS DR12 spectra\label{tab-SDSS}}
\tablehead{
Survey & ID   &   Spectroscopic & HJD & Exp. Time & $V_r$ & $\sigma V_r$ \\
            &      &    Survey                       & (d)  & (s)           & (\kms)    & (\kms) \\
}
\startdata
LSQ     & 662  & sdss  & 2451929.00553  & 4500.0  & 212.5 &  3.3 \\
LSQ     & 662  & sdss  & 2451660.75730  & 3600.0  & 221.9 &  3.5 \\
QUEST & 224  & sdss  & 2451609.90278  & 3600.0  &   6.2   &  3.9 \\
LSQ     & 1195 & segue1 & 2454581.74483  & 2700.0  &  -43.5 &  2.0 \\
LSQ     & 902    & boss     & 2455321.74094  & 3603.3   & -11.5  &  12.9 \\
\enddata
\tablecomments{Table~\ref{tab-SDSS} is published in its entirety in the electronic edition of The 
Astrophysical Journal, A portion is shown here for guidance regarding its form and content.}
\end{deluxetable*} 
 
\subsection{Spectroscopy}

The Sloan Digital Sky Survey (SDSS) contains a large number of stellar spectra from the SDSS-I, SDSS-II, SEGUE and BOSS surveys, 
which include RRLS. Presumably they were targeted as Blue Horizontal Branch (BHB)
stars in SDSS or quasar candidates, which may overlap in color with RRLS. 
These spectra have a wavelength coverage of 3800-9200{\AA } (3650-10400{\AA } for BOSS spectra), which for RRLS means that many Balmer lines
and several metallic lines (including Ca II H and K) are present. A potential problem about using spectra of RRLS
from SDSS is that there is the possibility that the final spectrum of a star is made of stacks of spectra taken 
non-consecutively, or even in different nights. Because the radial velocity of an RRLS changes by  up to $\sim 100$
{\kms } during the pulsation cycle, and the effective temperature and luminosity change by significant amounts as well, the composite of spectra taken at very different phases may yield a radial velocity that is far from the star's sytemic radial velocity by amount that is very hard to determine. To avoid this problem, the SDSS DR12 database{\footnote{\url https://www.sdss.org/dr12/}} was queried 
with the following restrictions: {\it (i)} all observations were done on a unique date (in practice, only one date listed in the 
keyword {\sl mjdList} in table {\sl PlateX}), {\it (ii)} there were no flags associated with the radial velocity ({\sl zwarning}$=0$), and
{\it (iii)} the total exposure time had to be less than 2 hours.
After these constraints, we found 423 usable spectra in DR12, several of which were multiple observations of the same 
star (Table~\ref{tab-SDSS}).

The first constraint ensures 
that several exposures over multiple nights were not combined into a single spectrum. 
The constraint on the total exposure times avoids excessive blurring due to the changing radial velocity. A 2-hour
exposure is equivalent to $15\%$ of the pulsation cycle of a typical ab-type RRLS (0.55d). The corresponding phase-span for the shorter periods 
c-type stars is much longer. However, the change in radial velocity for these stars is much lower than for the ab types, of the order of 30 \kms and thus, it is not expected that it will have a large effect on the final velocities. In practice, however,  the exposure times are rarely as large as two hours, and have a median of 1 hour.

In order to avoid the large discontinuity observed
in the radial velocity curve of ab-type RRLS at maximum light, we excluded any spectrum
taken near the maximum light phase of the pulsation cycle of the ab-type RRLS ($\phi>0.90$
or $\phi<0.10$). No restriction in phase was applied to the c-type stars since the discontinuity is not observed in these stars \citep[e.g.][]{vivas08}.
To implement this last restriction we calculated the phase
of observation using the HJD of the spectrum at mid-exposure and the ephemeris and periods
of the RRLS from the photometric catalogs. As said above, we gave preference to ephemeris provided by
LSQ in the case of stars in common among several catalogs. Heliocentric radial velocities were obtained from the keyword {\sl elodiervfinal} in Table
{\sl sppParams} in the case of the SDSS spectrograph, and from keyword {\sl elodieZ} in Table {\sl specObjAll} for BOSS spectra.

In addition to the SDSS spectra, several RRLS were observed by our team with the 4.1m SOAR 
(Cerro Pach\'on, Chile) and the 3.5m WIYN (Kitt Peak, 
USA) telescopes. 
At SOAR, we used the Goodman spectrograph with the 600 l/mm grating which yields 
a dispersion of 0.65 \AA/pixel and a wavelength coverage from 3500 to 6160 \AA. 
Six RRLS were observed with this telescope in February and March 2014. 
With WIYN, we used the Hydra multifiber positioner and grating 600@10.1 with the bench spectrograph 
to observe 26 additional RRLS (49 individual spectra) in the wavelength range 4000 to 6800 {\AA }  at 4.6{\AA } resolution.
Comparison lamps were 
obtained before and after each target spectrum with the Goodman spectrograph and after each spectrum with the WIYN bench spectrograph to ensure accurate wavelength solutions.
The data were reduced using standard IRAF packages. 

For the SOAR data, radial velocities were measured using 
Fourier cross-correlation with IRAF 
task fxcor and ELODIE radial velocity standards selected by \citet{duffau14} as templates. 
Following \citet{sesar13b} we used fxcor in two different ranges of wavelength, centered on H$\beta$
(4630-5000 \AA) and $H\gamma$ (4160-4630 \AA).  
For the WIYN data, better results were obtained by fitting profiles to individual lines and measuring their centers. In this case we also measured the $H\alpha$ line.
All velocities were corrected for the Earth's orbital motion, and they are reported in Tables~\ref{tab-SOAR} and ~\ref{tab-WIYN}. 

\floattable
\begin{deluxetable*}{rrrccccc}[htb!]
\tablecolumns{8}
\tablewidth{0pc}
\tablecaption{Heliocentric $H\beta$ and $H\gamma$ radial velocity for the RR Lyrae stars observed with the SOAR telescope\label{tab-SOAR}}
\tablehead{
Survey & ID   & HJD & Exp. Time & $V_r$ ($H\beta$) & $\sigma V_r$ ($H\beta$) & $V_r$ ($H\gamma$) & $\sigma V_r$ ($H\gamma$) \\
            &      & (d)  & (s)           & (\kms)    & (\kms) & (\kms) & (\kms) \\
}
\startdata
LSQ & 501 & 2456736.69039 & 2700 & 312.5 & 15.0  &  303.6 & 22.7 \\
LSQ & 525 &  2456736.64124 & 2700 &  324.6 & 13.0 & 362.3 & 25.7 \\
LSQ & 532 &  2456709.87263 &  900 &  41.8  & 15.5  & 44.4 & 15.6 \\
LSQ & 550 & 2456736.79199 & 3600 &  30.5 & 26.0 & -22.1 & 43.0 \\
LSQ & 691 & 2456736.74014 & 2700 & 74.4 & 10.1 & 89.8 & 17.9 \\
LSQ & 726 & 2456736.85050 & 2700 & 174.3 & 31.9 & 154.1 & 16.1 \\
\enddata
\end{deluxetable*}

\floattable
\begin{deluxetable*}{rrrcccc}[htb!]
\tablecolumns{7}
\tablewidth{0pc}
\tablecaption{Heliocentric $H\alpha$, $H\beta$ and $H\gamma$ radial velocity for the RR Lyrae stars observed with the WIYN telescope\label{tab-WIYN}}
\tablehead{
Survey & ID   & HJD & Exp. Time & $V_r$ ($H\alpha$) & $V_r$ ($H\beta$) & $V_r$ ($H\gamma$) \\
            &      & (d)  & (s)           & (\kms)    & (\kms) & (\kms) \\
}
\startdata
LSQ &  629 & 2455572.94156 &  2700   & -164.7 & -175.3 & -146.5 \\
LSQ &  629 & 2455574.95754 &  2700   & -124.4 & -131.3 &  -95.7 \\
LSQ &  606 & 2455572.94156 &  2700   &  203.9 &  184.8 &  194.1 \\
LSQ &  606 & 2455574.95754 &  2700   &  215.9 &  207.3 &  193.0 \\
LSQ &  611 & 2455572.94156 &  2700   &  207.5 &  192.2 &  236.9 \\
\enddata
\tablecomments{Table~\ref{tab-WIYN} is published in its entirety in the electronic edition of The 
Astrophysical Journal, A portion is shown here for guidance regarding its form and content.}
\end{deluxetable*}

To complement our sample, we added the velocity measurements of RRLS 
in the region which are already in the literature. This includes data from
\citet{duffau14} (79 stars with distances $<22$ kpc), \citet{vivas05} (10 stars in the
Sgr stream located at distances in the range 40-55 kpc), as well as 5 SEKBO RRLS \citep{prior09} which are located 
in this region of the sky. 
Finally, new observations for 14 distant QUEST stars (distances $>30$ kpc) were taken from a spectroscopic study of the Sagittarius tidal tails
(Duffau et al. in preparation). 
These stars were observed with Gemini South, Magellan and VLT (FORS2) telescopes with instrumental setups
similar to others used in this work. The reduction techniques and the method of measurement of radial velocities were similar
to the ones used by \citet{duffau14}.

\begin{figure}[htb!]
\plotone{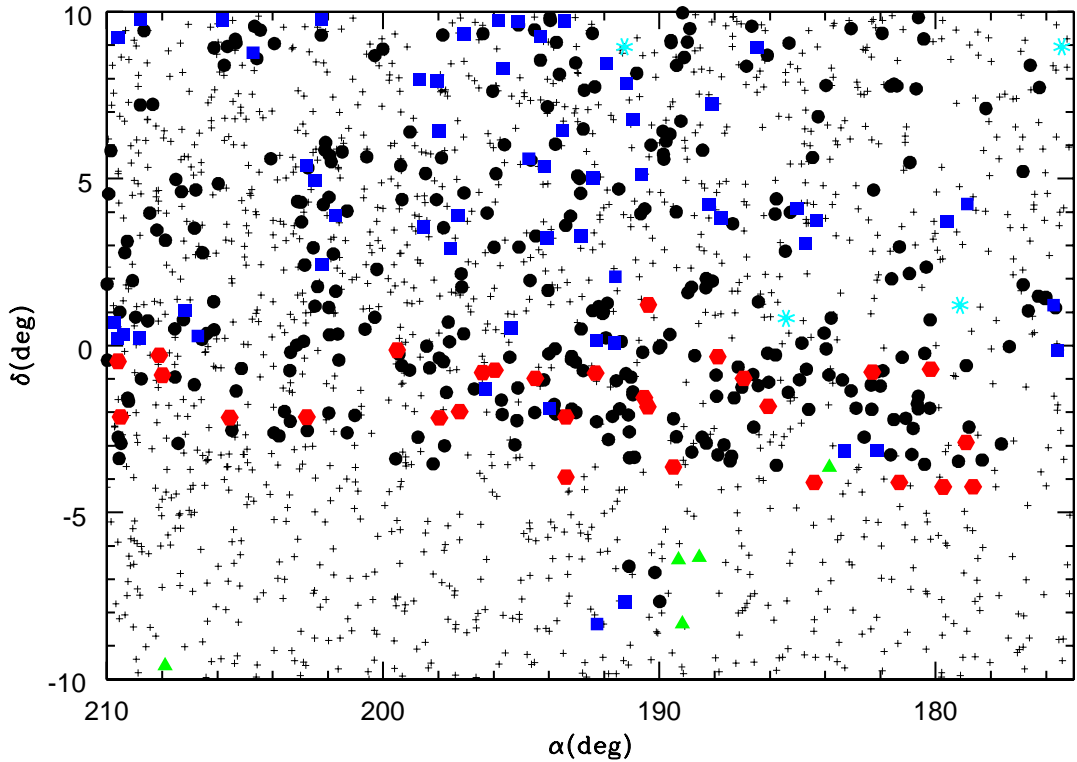}
\caption{Distribution in the sky of the known RR Lyrae stars (RRLS) in the Virgo overdensity region (small crosses). Those stars with
spectroscopic observations are highlighted with larger symbols. The different symbols
refer to the photometric catalog of RRLS (LSQ, black circles; QUEST, red hexagons; CRTS, blue squares;
LINEAR, cyan asterisks; SEKBO, green triangles.}.
\label{fig-sample}
\end{figure}

There are stars in common between the different spectroscopic datasets described above. Their systemic velocities were determined by fitting a radial velocity curve to the available observations. 
In total, we used 681 individual spectra of 472 RRLS. The distribution in the sky of our final sample is shown in 
Figure~\ref{fig-sample}. This spectroscopic sample corresponds to $\sim 30\%$ of the known RRLS in this region of the sky.

\section{Systemic Radial Velocities}

To obtain the systemic radial velocity of RRLS it is necessary to subtract the 
velocity due to the pulsation of the star. In this work, we adopted the radial velocity curves 
provided in \citet{sesar12} which are based on an extensive set of measurements for six 
type ab RRLS. The advantage of using these templates is twofold. First, there are 
separate templates for the Balmer lines H$\alpha$, H$\beta$ and $H\gamma$ and 
for metallic lines. We can then choose the appropriate template(s) for the wavelength range of
a particular spectrum. Second, \citet{sesar12} provides a calibration between the 
amplitude of the light curve and the amplitude of each one of the radial velocity curve templates.
The use of these templates and calibration represent an improvement for the determination 
of the systemic velocities 
compared to the traditional use of the model of a  single star (X Arietis), which was constructed 
with measurements of the H$\gamma$ line and had a fixed amplitude which was assumed
to hold for any RRLS \citep[see for example][]{layden95,vivas05,vivas08,
prior09,duffau14}. 
For uniformity, we used these templates to re-calculate the systemic velocities of all stars in our sample,
including the ones previously reported in the literature.

For type ab stars we followed the following steps in order to derive the systemic radial velocities:

\begin{itemize}
\item Depending on the wavelength range of each particular spectrum, we selected which template(s) were to be used for correcting the pulsation velocity. For example, 
SDSS velocities are measured by Fourier cross-correlations using the complete spectral range, which contain the three Balmer lines considered above and a wealth of metallic lines. Consequently, all four templates were used on these spectra. On the other hand, for the SOAR spectra we only used the H$\beta$ and H$\gamma$
templates, while for the WIYN data, the three Balmer line templates were used. 

\item Each template was scaled by the amplitude of the light curve of the RRLS according to equations 2-5 in \citet{sesar12}. 

\item Having calculated the mean phase of observation of a spectrum, we interpolated 
in each one of the selected radial velocity templates to obtain a pulsational velocity at that phase

\item The pulsation velocity was subtracted from the measured velocity to give the systemic velocity for each template. In the case of SDSS
spectra, the measured velocity is the one reported in Table~\ref{tab-SDSS}, while for the SOAR and WIYN spectrum is the velocity measured
for the specific Balmer line of the template.

\item  The results from the different templates were weighted by their errors and then averaged to yield our final systemic velocity. 
The resulting systemic velocities derived from each template were not very different. For example, for the SDSS data, the standard deviation 
of the four velocities obtained from each template averaged 15 \kms.

\item If more than one spectrum for the same star was available, the systemic velocities were averaged.

\item The errors of systemic velocities were estimated following \citet{sesar12} and \citet{sesar13b}, which included the error in the
velocity measurement and the error of the template at the phase of observation.

\end{itemize}

\begin{figure}[htb!]
\plotone{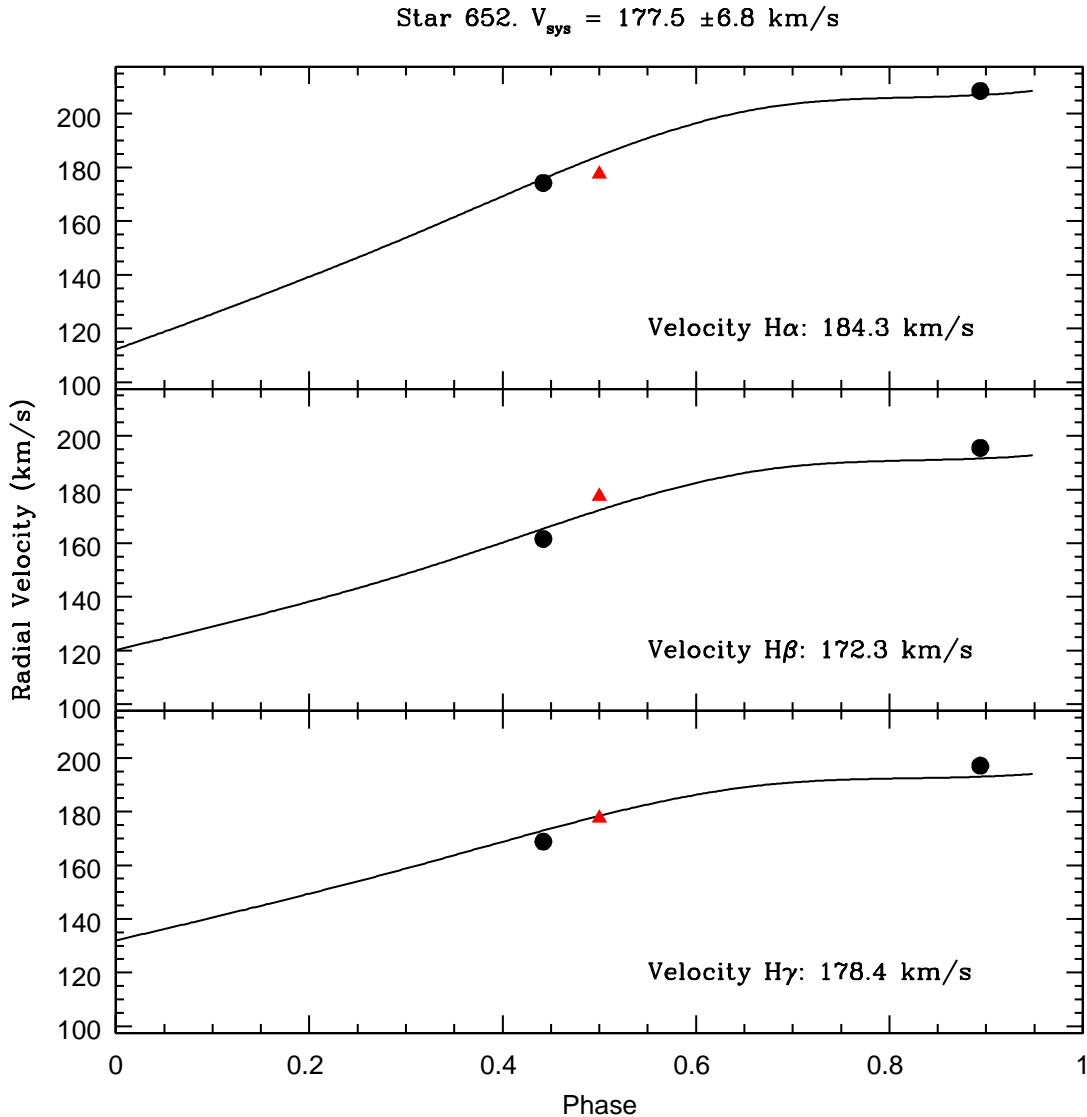}
\caption{Radial velocity curve templates fitted to star LSQ 652 which was observed with the WIYN 
telescope. Each panel shows, as solid circles, the $H\alpha$, $H\beta$ ans $H\gamma$ measurements for two spectra
taken at phases 0.44 and 0.89. The solid lines show the fitted templates which have been scaled by
the amplitude of the light curve of this star. As reference, the red triangle indicates the final systemic velocity obtained 
after averaging (with weights) the results from each template fitting.}
\label{fig-rv652}
\end{figure}

Figure~\ref{fig-rv652} illustrates the radial velocity curve
fitting for star LSQ 652 ($<V>=17.0$) which was observed with the WIYN telescope at two different phases.

The same approach described above was followed to recalculate the systemic velocities in the catalog of \citet{duffau14} 
since they used X Arietis as the radial velocity template.  A comparison of the present results with those of \citet{duffau14} show 
that the mean difference in radial velocity 
for the stars in common is only 1 \kms, with a standard deviation of 19 \kms. The differences are partly due to the use
of the new radial velocity templates, but a significant contribution comes from our use of updated ephemerides in this work 
(those from the LSQ survey). Indeed, if we use the same ephemerides as \citeauthor{duffau14} to fit the new radial velocity curves, the standard deviation of the
velocity differences decreases to only 9 \kms.  This suggests that the results of previous investigations that used only X Arietis as a template are not seriously biased by the choice of template. 

As mentioned before, we included 5 RRLS with kinematical data from \citet{prior09} which lie in the range
$183\degr < \alpha < 208\degr$ and $-3\degr < \delta < -10\degr$. 
These stars are important because they 
extend our coverage toward southern declinations.
The systemic radial velocities of these stars were calculated using the star X Arietis as template, and we were unable to
re-calculate the velocity using the new templates because there was no information on
the individual spectra for each star in \citet{prior09}. The above comparison with the results of \citet{duffau14} suggests that 
that the velocities given by \citet{prior09} may be used without any correction.

RRLS of the type c have smaller variations in the radial 
velocities during the pulsation cycle. For these stars we used the template given in
\citet{vivas08} and \citet{duffau14} to derive
the systemic radial velocities.

The final sample of RRLS with
radial velocities in the Virgo region consists of 412 stars, which is roughly 6 times larger than the 
sample in \citet{duffau14}, and it covers a much larger area and range of distances from the Sun ($\sim 4 - 75$ kpc).
Final results are shown in Table~\ref{tab-results}. The table contains the photometric survey and the identification number of the
RRLS in that survey, coordinates, type of RR Lyrae, mean V magnitude, period of pulsation, amplitude in the V band, distance
from the Sun, number of useful spectra used to derive the systemic radial velocity, heliocentric radial velocity, radial 
velocity in the galactic standard of rest frame\footnote{Assuming a solar motion of $(v_U, v_V, v_W) = (10.0, 5.2, 7.2)$ {\kms } 
and $v_{LSR} = 220$ {\kms } \citep{binney98}.} and associated error, and finally, the group number to which the star belongs according to
our analysis in the next section.

\floattable
\begin{deluxetable*}{rrccrcccrcrrrc}[htb!]
\tabletypesize{\scriptsize}
\tablecolumns{14}
\tablewidth{0pc}
\tablecaption{Position, light curve parameters, distance and radial velocity for the RR Lyrae stars\label{tab-results}}
\tablehead{
Survey & ID   &   RA(2000.0) &  DEC(2000.0) &  Type &  Mean V & Period & Amp V & 
Distance  & N$_{\rm spec}$ & $V_h$  & $V_{gsr}$ & $\sigma V_h$ & Group \\
           &        &  (deg)            & (deg)              &           &              & (d)      & (mag)          &
 (kpc)      &                            & (km/s)  & (km/s)      & (km/s)              &   \\
}
\startdata
   LSQ &                     381 & 178.18130 &   7.10520 & ab & 16.21 & 0.48523 &  1.29 & 13.3 & 1 &  254 &  167 & 16 & \\
   LSQ &                     384 & 178.31850 &  -3.42868 & ab & 16.80 & 0.56959 &  0.81 & 17.2 & 1 &  182 &   64 & 17 & \\
 QUEST &                     717 & 178.64810 &  -4.23051 & ab & 15.63 & 0.66514 &  1.01 & 10.0 & 2 &  229 &  110 & 14 & 9 \\
   LSQ &                     393 & 178.79402 &  -2.44436 & ab & 16.84 & 0.53839 &  0.95 & 17.6 & 1 &  206 &   92 & 18 & \\
   CRTS &    CSS\_J115524.9+041414 & 178.85396 &   4.23725 & ab & 18.20 & 0.68168 &  1.02 & 31.8 & 1 &   50 &  -45 & 17 & \\
\enddata
\tablecomments{Table~\ref{tab-results} is published in its entirety in the electronic edition of The 
Astrophysical Journal, A portion is shown here for guidance regarding its form and content.}
\end{deluxetable*}
 
\begin{figure}[htb!]
\plotone{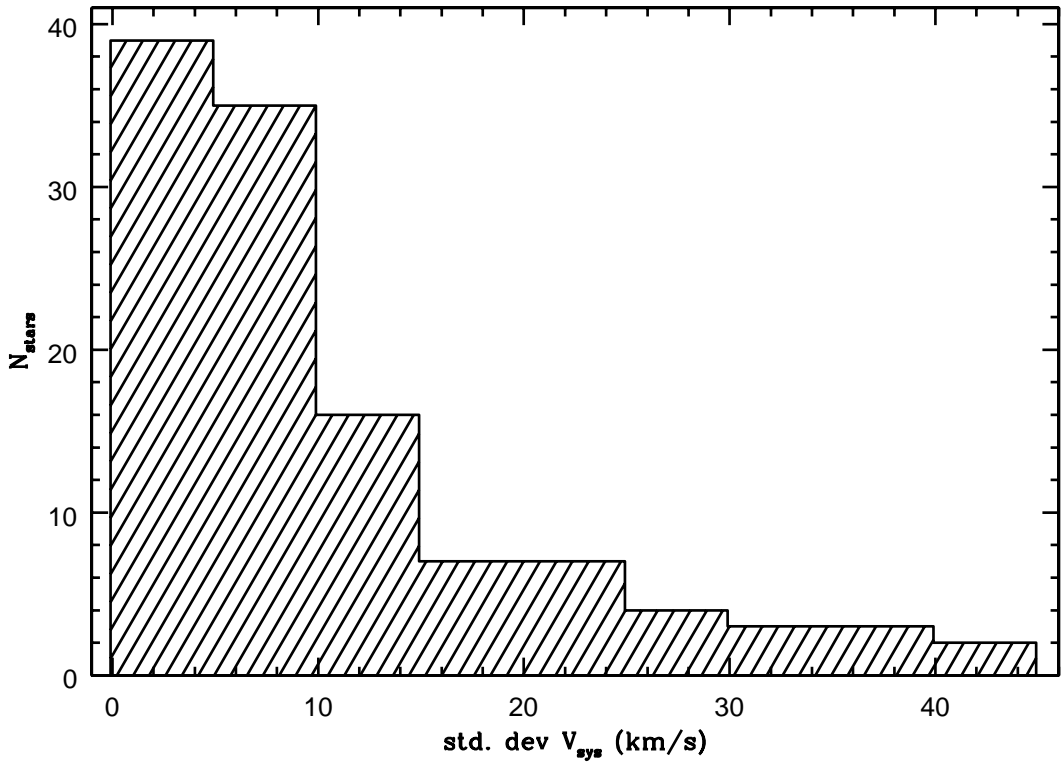}
\caption{Standard deviation of the systemic velocities $V_{sys}$ obtained for stars with multiple spectra available.}
\label{fig-multiple}
\end{figure}

About 1/3 of the sample (116 stars) have more than one spectrum taken at different phases. In several 
cases, the spectra were taken with different instruments. In order to check the consistency of our
method we calculated the standard deviation of the systemic velocities obtained by stars with more than 
one spectrum available (Figure~\ref{fig-multiple}). The median value is only 8 {\kms } which is encouraging that we are correcting appropriately the pulsation velocity of the RRLS and that there are no large systematic differences 
between 
different instrumental setups. Only a few stars had standard deviation in their velocities of the order of 
25-45 \kms. Every time the standard deviation from multiple 
spectra was larger than the error in the systemic velocity (which includes both the observational and the template error), 
we adopted the former as the final error.

The average error in radial velocity for all the stars in the sample is 13.8 \kms. There is no significant trend of the radial velocity errors with distance to 
the star, and thus, we adopted this mean error for the analysis described next. In distance, we assumed
a 7\% error, following \citet{zinn14}.

Star LSQ 394 (QUEST 167) requires further explanation. This star was used by \citet{casetti09} to determine a preliminary orbit for the VSS, which was 
later confirmed by \citet{carlin12} using stars consistent with the color-magnitude diagram of the VSS. The radial velocity calculated by \citeauthor{casetti09} used the same 
SDSS spectra used here and the ephemerides given by the QUEST survey, which was based on a light curve with 29 observations. LSQ has a superior lightcurve based on 158 epochs, which we have used to re-derive the radial velocity.  We obtained $V_{gsr} = 180$ \kms, which is  higher than the value
of 134 {\kms } reported by \citet{casetti09}, but still consistent, within error to the mean velocity of proper-motion VSS stars in \citet{carlin12}.
The LSQ distance, however, is compromised because the photometric calibration for this star was flagged as being poor, which means it can produce distance errors of $\sim 12\%$ \citep{zinn14}. Consequently, we adopted for this star the distance determined by the QUEST survey, 16.9 kpc.
 
\section{Kinematical Groups \label{sec-groups}}

\subsection{Method}

We searched for kinematical grouping in the data using the method developed
in our previous study of the region \citep{duffau14} with some modifications, which are explained below. This new implementation of the method makes
it more robust in a sample covering a large range of distances, as encountered here.
The method is basically a friends-of-friends algorithm which selects pairs of stars which are located in 
nearby (3D) positions in the sky and share a similar velocity, or formally, 
stars that have a four-distance value less than a critical value, or $4d< \epsilon$ \citep{starkenburg09}. 
Pairs with common stars (friends-of-friends) form a group. 
The $4d$ parameter is defined as:

\begin{equation}
\centering
(4d_{ij})^2 = \omega_{3d_{ij}} (3d_{ij})^2 + \omega_v (v_i-v_j)^2 
\label{eq-old4d}
\end{equation}

\noindent
where ($3d_{ij}$) is the mean physical separation between stars $i$ and $j$, which can be calculated using their
Galactocentric cartesian coordinates (assuming $R_\odot=8$ kpc):

\begin{equation}
\centering
3d_{ij} = \sqrt{(x_i-x_j)^2 + (y_i-y_j)^2 + (z_i-z_j)^2}
\end{equation}

As defined in \citet{starkenburg09} and \citet{duffau14}, the weights  $\omega_v$
and $\omega_{3d_{ij}}$ in equation~\ref{eq-old4d} depend on both the maximum possible range for those quantities
and the observational errors. In the case of $\omega_{3d_{ij}}$, the observational error appeared
in the form of a relative error ($\sigma d/d$). Using those weights, we searched for sub-structures
in \citet{duffau14} using $\epsilon=0.045$, which implied a separation between
stars in a pair $\lesssim 25$ \kms and $\lesssim 3$ kpc, in velocity and separation in the sky respectively. 
The maximum of 3 kpc for the separation of stars in a pair was adequate given the distance for the RR
Lyrae stars used in \citet{duffau14} which were all located closer than 23 kpc from the Sun (hence, with errors in distance
$<1.6$ kpc). In the present investigation, 
we have stars at $>50$ kpc, which implies errors in distance $> 3.5$ kpc. Obviously, we do not want to limit the search to pairs of stars with separations 
less than the errors in their distances.

Thus, in this implementation of the algorithm 
we modified the $4d$ equation in order to use the absolute error in distance (instead of its relative error), and
we normalized it in such a way that a value of $\epsilon$ corresponds to a maximum separation of $N$ times the error in both distance and velocity.
Weights have the form:

\begin{equation}
\omega_v = \frac{a}{(v_{max})^2}; \;\;\;\;\; \omega_{3d_{ij}} = \frac{b}{((3d)_{max})^2}
\end{equation}

\noindent
where the denominators are the maximum possible range for stars in our sample, which are needed in order to make $4d$ a dimensionless variable.  The values $a$ and $b$ are determined by the following procedure.

When two stars in a pair have the same position ($3d_{ij}=0$) they differ only in velocity. The value of $\epsilon$ sets what is the maximum possible difference in velocity for the stars in a pair. From Eq (\ref{eq-old4d}):

\begin{equation}
4d_{ij} = \sqrt{\omega_v ((v_i - v_j)_{max})^2} = \epsilon
\end{equation}

We request that the maximum difference in velocity scales with the error in velocity, that is, $(v_i-v_j)_{max}  = N \sigma_v$. 
Thus,

\begin{equation}
a = \left( \frac{\epsilon}{N} \right)^2 \left( \frac{v_{max}}{\sigma v} \right)^2
\label{eq-a}
\end{equation}

Similarly, a maximum separation of two paired stars in the sky ($Sep_{max}$) is achieved if their velocities are the same ($v_i-v_j=0$),
which gives

\begin{equation}
4d_{ij} =  \sqrt{ \omega_{3d_{ij}} (Sep_{max})^2} = \epsilon
\end{equation}

And in this case we require that  $Sep_{max} = N \sigma d_{ij}$ and obtain:

\begin{equation}
b = \left( \frac{\epsilon}{N} \right)^2 \left( \frac{(3d)_{max}}{\sigma d_{ij}} \right)^2
\label{eq-b}
\end{equation}

Combining (\ref{eq-a}) and (\ref{eq-b}), we obtain:

\begin{equation}
a \left(\frac{\sigma v}{v_{max}} \right)^2 = b \left(\frac{\sigma d}{(3d)_{max}} \right)^2
\end{equation}

Choosing (arbitrarily) $a=1$, we can solve for $b$ and obtain the following relationship for the 4-distance::

\begin{equation}
\centering
(4d_{ij})^2 =  \left(\frac{1}{v_{max}} \right)^2  \left( \frac{\sigma v}{\sigma d_{ij}} \right)^2 (3d_{ij})^2 + \left( \frac{1}{v_{max}} \right)^2(v_i-v_j)^2  < \epsilon
\label{eq-new4d}
\end{equation}

\noindent
and $\epsilon = N \sigma v / v_{max}$.

Here, $\sigma v$ is the mean error in radial velocity of the full sample, while $\sigma d_{ij}$ is the mean error in distance of the stars in a pair.
Using $N=2.0$, a pair of stars will have a maximum possible difference in velocity of 2 times the mean error in the radial velocity, which translates
to 27.6 \kms given the mean error of our sample. Unless the stars are located in exactly the same spot, the velocity difference between them will be less than this value. Similarly, a pair can have a maximum separation of 2 times the error in the mean distance of the pair, which equals 1.4 kpc at 10 kpc or 7 kpc at 50 kpc, if the stars have exactly the same radial velocity. 

\begin{figure*}[htb!]
\plotone{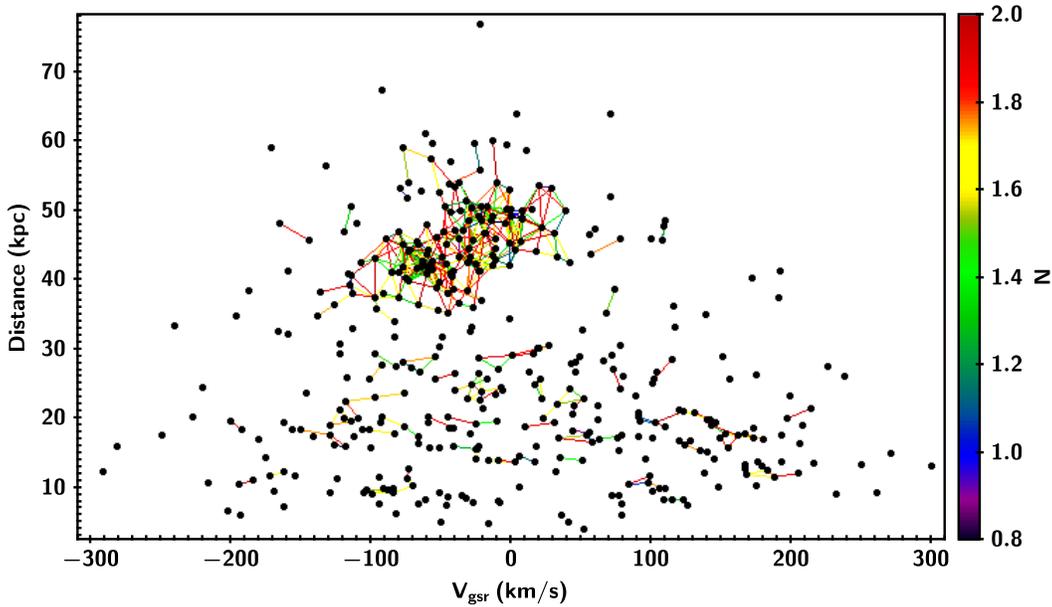}
\caption{Heliocentric distance versus radial velocity (in the galactic standard of rest frame) for all the RRLS studied in this work. Pairs of stars selected with $N<2.0$ are connected by solid lines whose color scale with $N$. }
\label{fig-pairs}
\end{figure*}

\begin{figure*}[htb!]
\plotone{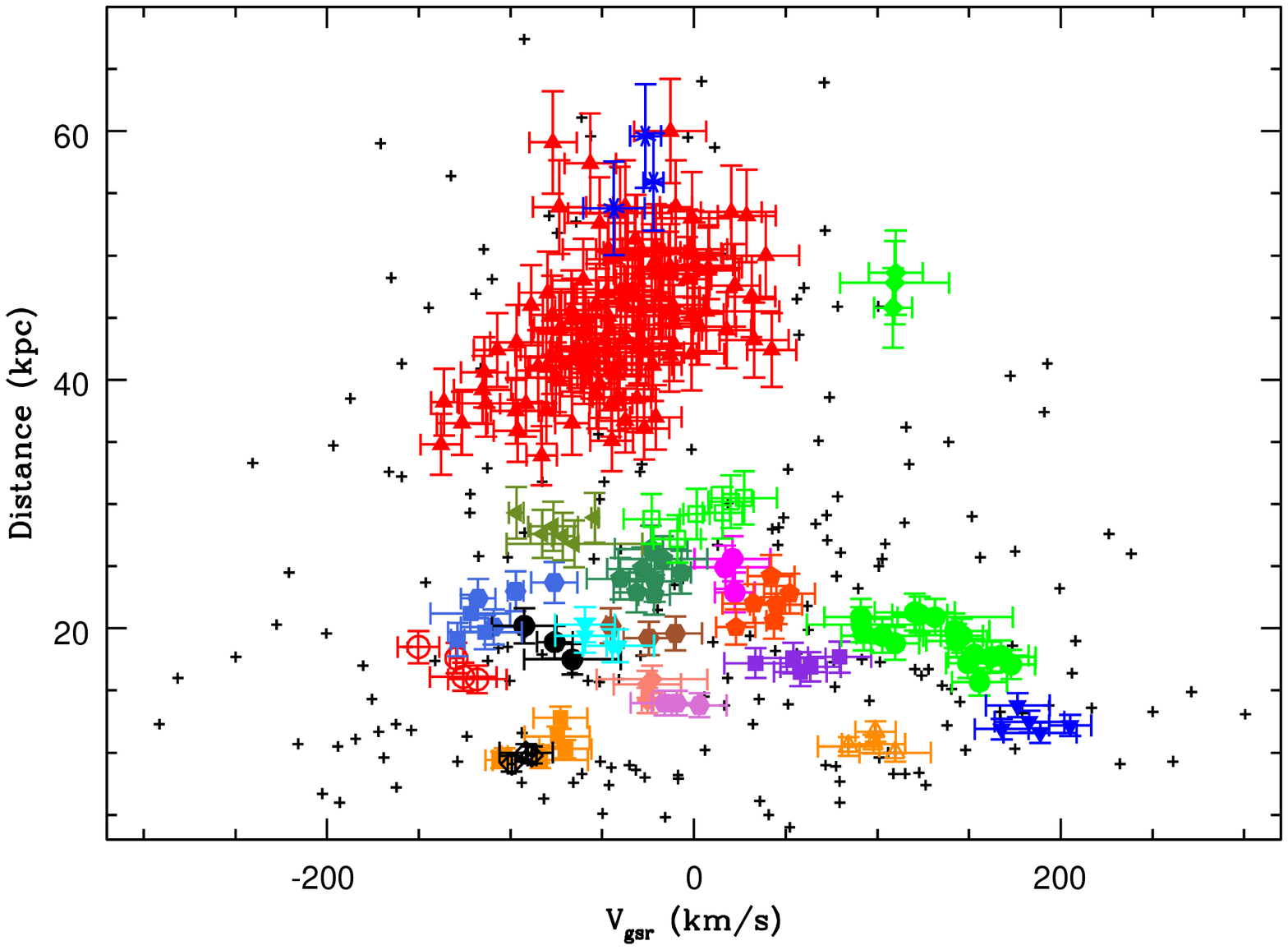}
\caption{Heliocentric distance versus radial velocity (in the galactic standard of rest frame) for all the RRLS studied in this work. Colored large symbols indicate different groups whose stars have pairs of stars with
4-distance $4d<2.0\sigma v$.}
\label{fig-allgroups}
\end{figure*}

Applying equation~\ref{eq-new4d} to our sample, we found 558 pairs of stars (Figure~\ref{fig-pairs}) and 22 groups (Figure~\ref{fig-allgroups}). 
Most of the groups of RRLS have a small number of members ($\leq8$) but two of them have particularly large number of stars, 
one with 18 members at a distance 
of $\sim 19$ kpc and another one with 113 members at $\sim 45$ kpc.

It is likely that some of these groups, particularly the ones with a small number of members, are produced by random variations in a smooth distribution of stars. In order to
quantify how frequently these cases may occur, we simulated a large number of random samples of halo stars, which were
used as input in our group-finding algorithm. We then quantified how likely is to form random groups as a function of number
of members, mean radial velocity, mean distance, velocity dispersion and distance dispersion. With this procedure, we were
able to identify which of the RRLS groups are unlikely be simply formed by random in a smooth distribution of stars.

We made 10,000 simulated samples with the following recipe.  First, we estimated how many RRLS are expected in this area of the sky in the range of 
distances between 4 and 70 kpc by integrating the RRLS number density profile provided in \citet{zinn14}.  For simplicity the area
of observations was approximated to a rectangular region with limits $175\degr < \alpha < 210\degr$ and $-4\degr < \delta < 10\degr$, which is where most
of our spectroscopic sample is found (see Figure~\ref{fig-sample}).
This integration yielded 767 RRLS in this volume of the halo. Because the completeness of the LSQ
survey (which was used to derive that profile) is estimated to be 70\%, we divided this number by 0.7 to obtain the true expected number of RRLS. However,
our spectroscopic sample is not complete; only $30\%$ of the real RRLS have at least one spectrum available. 
Taking into account this factor, 
we simulated samples of stars having $N_{\rm stars} \pm \sqrt{N_{\rm stars}}$, where $N_{\rm stars}=329$.
Although the spectroscopic completeness varies some with location on the sky because the different datasets were combined,  we assumed for simplicity
uniform completeness in our simulations. 

To each star in the simulated sample, we assigned a random position, heliocentric distance and radial velocity. The positions were randomly assigned within the rectangular region defined above.  
Distances were randomly drawn from a distribution that is based on the number density radial profile of RRLS \citep{zinn14}, in the
range galactocentric distances ($R_{gc}$) from $\sim 7$ to 90 kpc. This profile has an flattening of $c/a=0.7$ and a double power law with a steeper slope beyond $\sim 30$ kpc. 
A $7\%$ error, similar to the observational error of the real RRLS,  was added to the resulting distance. Radial velocities ($V_{gsr}$) 
were randomly assigned following a Gaussian distribution with mean 0 \kms and standard deviation which was a function of the distance 
from the galactic center, following the velocity dispersion profile given by \citet{brown10} for BHB stars in the halo. 
Brown et al.'s profile is
defined from $R_{gc} = 15$ to 80 kpc. We assumed that the velocity dispersion at 15 kpc (103 \kms) also holds to the inner limit of our sample ($R_{gc} \sim 7 $ kpc). The assumed velocity dispersion at a 
given distance took into account the errors in the profile derived by \citeauthor{brown10}.
Noise with the typical observational error of 13.8 {\kms } was added to the assigned radial velocity.

\begin{figure}[ht]
\plotone{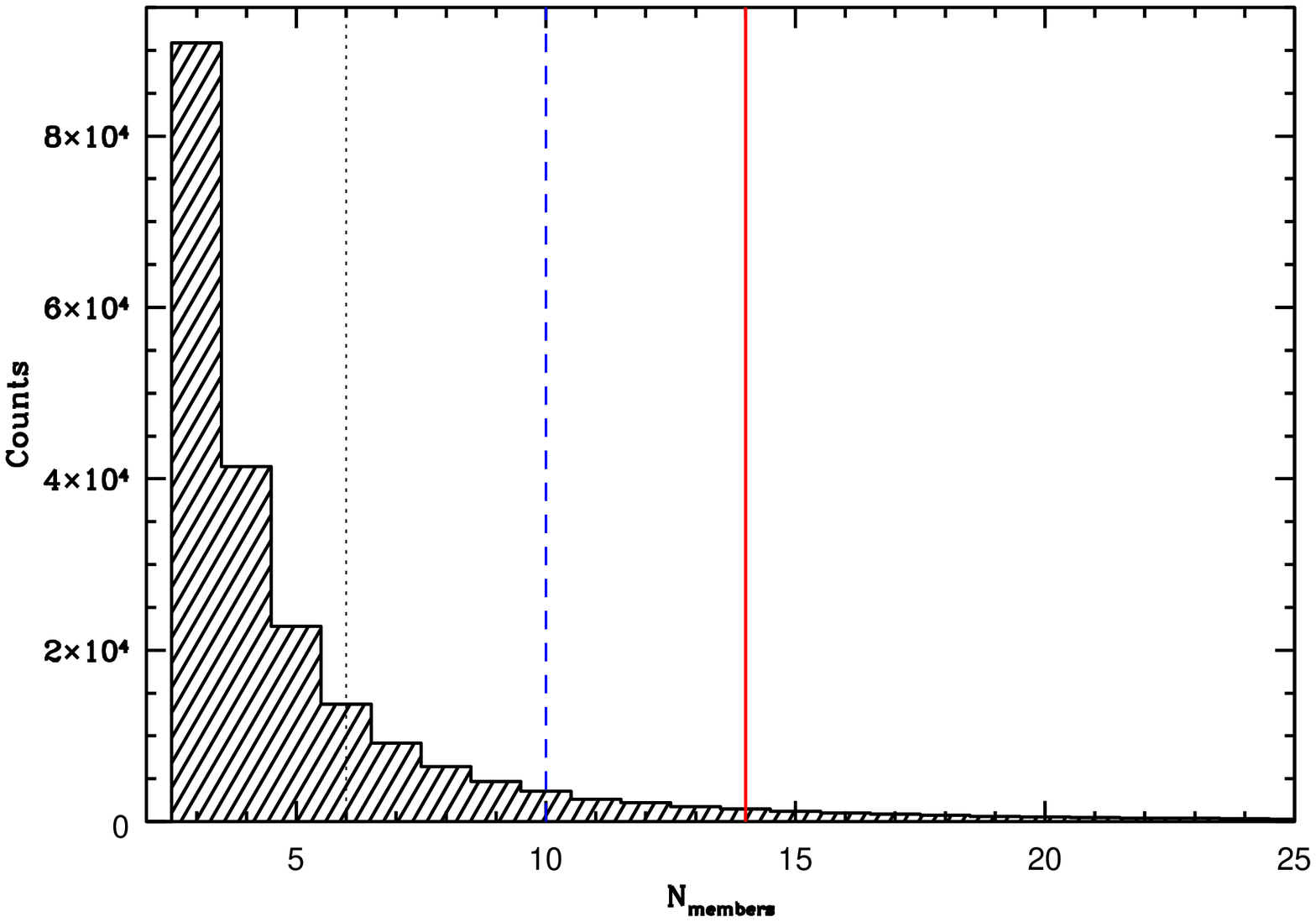}
\caption{Number of members in 210,298 fake groups detected with $4d<2.0\sigma v$ in all of the simulated samples.
The vertical lines indicate the 75th percentile (dotted black), 90th percentile (dashed blue) 
and 95th percentile (solid red).}
\label{fig-Nmembers}
\end{figure}

\subsection{Results}

As expected, our algorithm detected pairs and groups in the simulated samples. 
The mean number of pairs found in the simulated samples is 148, which is significantly lower than the 558 pairs found among the RRLS. 
This suggests this region of the sky contains sub-structures.
The number of groups, however, are about the same, 21 on average in the simulated data and 22 in the real data.  
Evidently, many of the pairs in the real data originate in only a few sub-structures.
The real data contain indeed two relatively large structures, with 113 and 18 members, in which our algorithm connected many pairs of stars.  In contrast, the simulations produced mostly groups with small numbers of members (Figure~\ref{fig-Nmembers}).  The median number of
members among
the 210,298 groups formed in the 10,000 simulations is 4 stars. The 95th percentile of the distribution of the number of members is 14, and the largest group formed in all 10,000 simulations has 83 members.  Based on size alone, the most populous groups in the RRLS sample are very likely to be real, and in fact they are due to the previously well-documented stream of stars from the Sgr dwarf sphoidal (dSph) galaxy and the Virgo Stellar Stream (VSS).  

\begin{figure}[ht]
\plotone{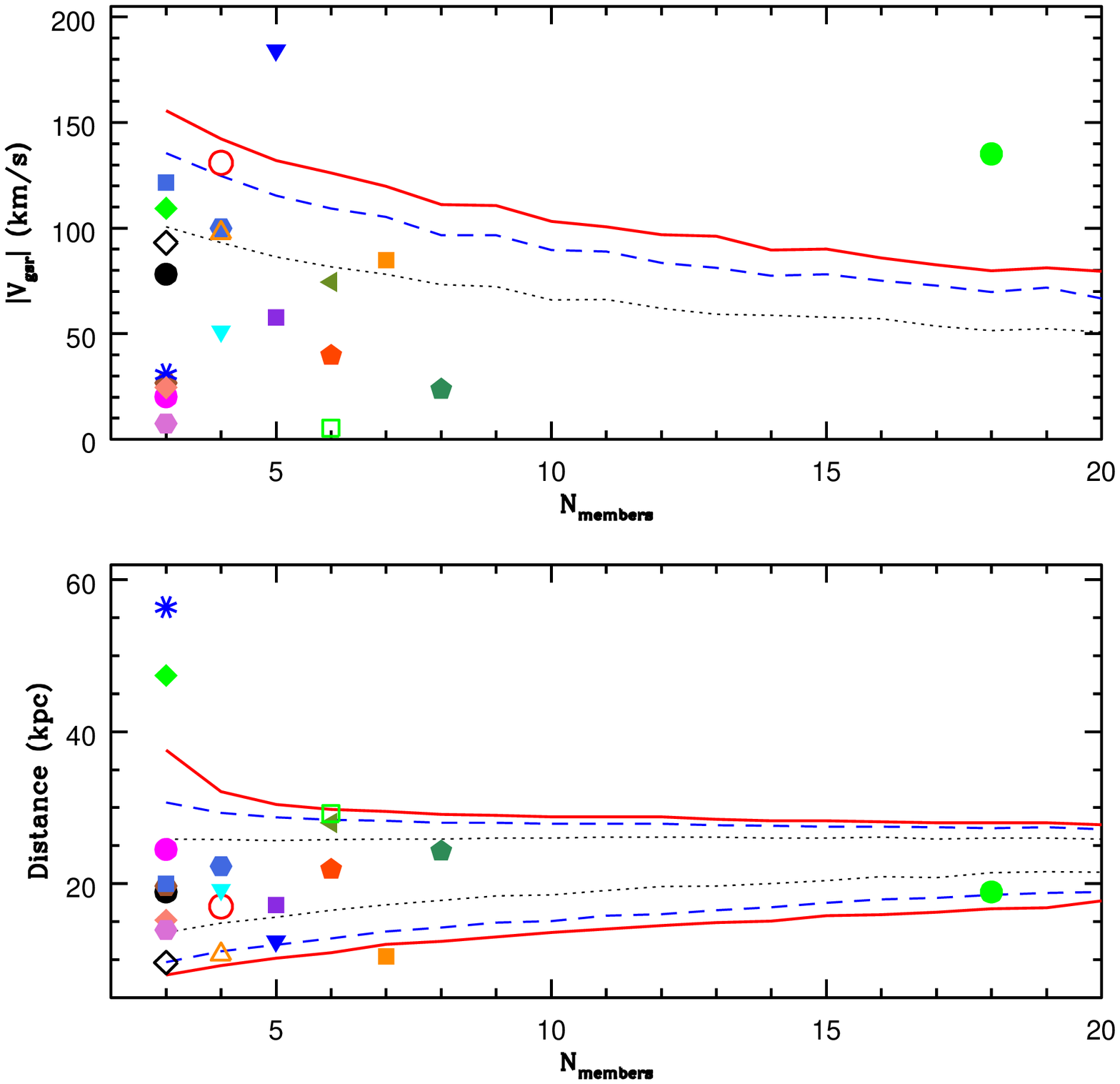}
\caption{For each group of RR Lyrae stars (colored symbols) we show the absolute value of the mean radial velocity (top) and  mean heliocentric distance (bottom)
as a function of the number of members in each group. 
The color and type of the symbol for the RRLS groups are the same as the ones use in
Figure~\ref{fig-allgroups}. The lines represent the 5th, 95th percentile (solid red line), 10th, 90th percentile (dashed blue line) and 
25th, 75th percentile (dotted black line) of the distribution of those parameters in the fake groups found in our simulations. 
Only the 95th, 90th and 75th are shown for the top panel.
Some groups of RRLS are located well above of the 95th percentile line in velocity and distance, indicating very few
random groups are formed with those parameters.
}
\label{fig-percentile}
\end{figure}

It is much less clear that any of the smaller groups that we have detected in the RRLS sample (Figure~\ref{fig-allgroups}) are in fact real halo 
substructures. To examine this question, we compare in Figure~\ref{fig-percentile} values of $|V_{gsr}|$ and distance from the Sun of the groups in the 
RRLS sample with the distributions of these quantities among the fake groups produced by the simulations.  The lines in diagrams in 
Figure~\ref{fig-percentile} indicate the values of $|V_{gsr}|$ and Distance corresponding to the 5th, 10th, 25th, 75th, 90th, and 95th percentiles of the distribution of fake groups in the 10,000 simulations as a function of the group size $N_{\rm members}$.
Any group whose velocity and/or distance is far from the mean values of these quantities in fake groups of the same 
$N_{\rm members}$ is likely real and will be flagged as significant. The group of RRLS with 113 members is not plotted because none of the fake groups was so populated.  

To illustrate the usefulness of Figure~\ref{fig-percentile}, consider the group of RRLS plotted as blue upside down triangle.  It has only 5 
members, and groups of that size are frequently produced in the simulations (Figure~\ref{fig-Nmembers}).  But the mean $|Vgsr|$ of this group is 184 \kms, and comparable values are very rarely seen in 5-member groups in the simulations (the upper 95th percentile is $\sim 130$ \kms) because this 
value is in the tail of the velocity distribution of halo stars.  This strongly suggests that this group of 5 RRLS is part of a real halo substructure.  

In Table~\ref{tab-groups}, we identify as groups of high significance the ones that have either more than 14 members and/or values of $|V_{gsr}|$ and distance that lie outside the regions in 
Figure~\ref{fig-percentile} enclosed by the solid lines (5th and 95th percentile).  Groups of lower significance are identified if they fall outside the regions enclosed by the 
dashed lines in Figure~\ref{fig-percentile} (10th and 90th percentile).  
The parameters that caused a group to be identified as significant are printed in boldface.     
RRLS that are members of significant groups are identified by their group number in the last column of Table~\ref{tab-results}.        

\floattable
\begin{deluxetable*}{cccccl}[htb!]
\tabletypesize{\footnotesize}
\tablecolumns{5}
\tablewidth{0pc}
\tablecaption{Significant groups of RR Lyrae stars\label{tab-groups}}
\tablehead{
Group & $\langle V_{gsr} \rangle$ & $\langle d \rangle$ & $N_{\rm members}$ & Symbol/color \\
           & (\kms)                               & (kpc)                       &                                  &                      \\
}
\startdata
\cutinhead{High significance groups}
1 & -41.3            & {\bf 44.7}    & {\bf 113}  & triangles/red \\
2 & {\bf 135.2}  & 18.9             & {\bf 18}    & circles/green \\
3 & -84.8          & {\bf 10.4}      & 7              & squares/orange \\
4 & {\bf 184.2}  & 12.4             & 5              & upside-down triangles/blue \\
5 & 109.4           & {\bf 47.4}    & 3              & diamonds/green \\
6 & -30.7            & {\bf 56.4}    & 3              & asterisks/blue \\
\cutinhead{Low significance groups}
7 & 5.4               & {\bf 29.2}   & 6               & open squares/green \\
8 & {\bf -130.9} & 17.0            & 4              & open circles/red \\
9 & 97.7            & {\bf 10.7}     & 4              & open triangles/orange \\
10 & -93.1         & {\bf 9.6}       & 3              & open diamonds/black \\
\enddata
\end{deluxetable*}

\begin{figure}[t]
\centering
\includegraphics[angle=270,width=0.7\textwidth]{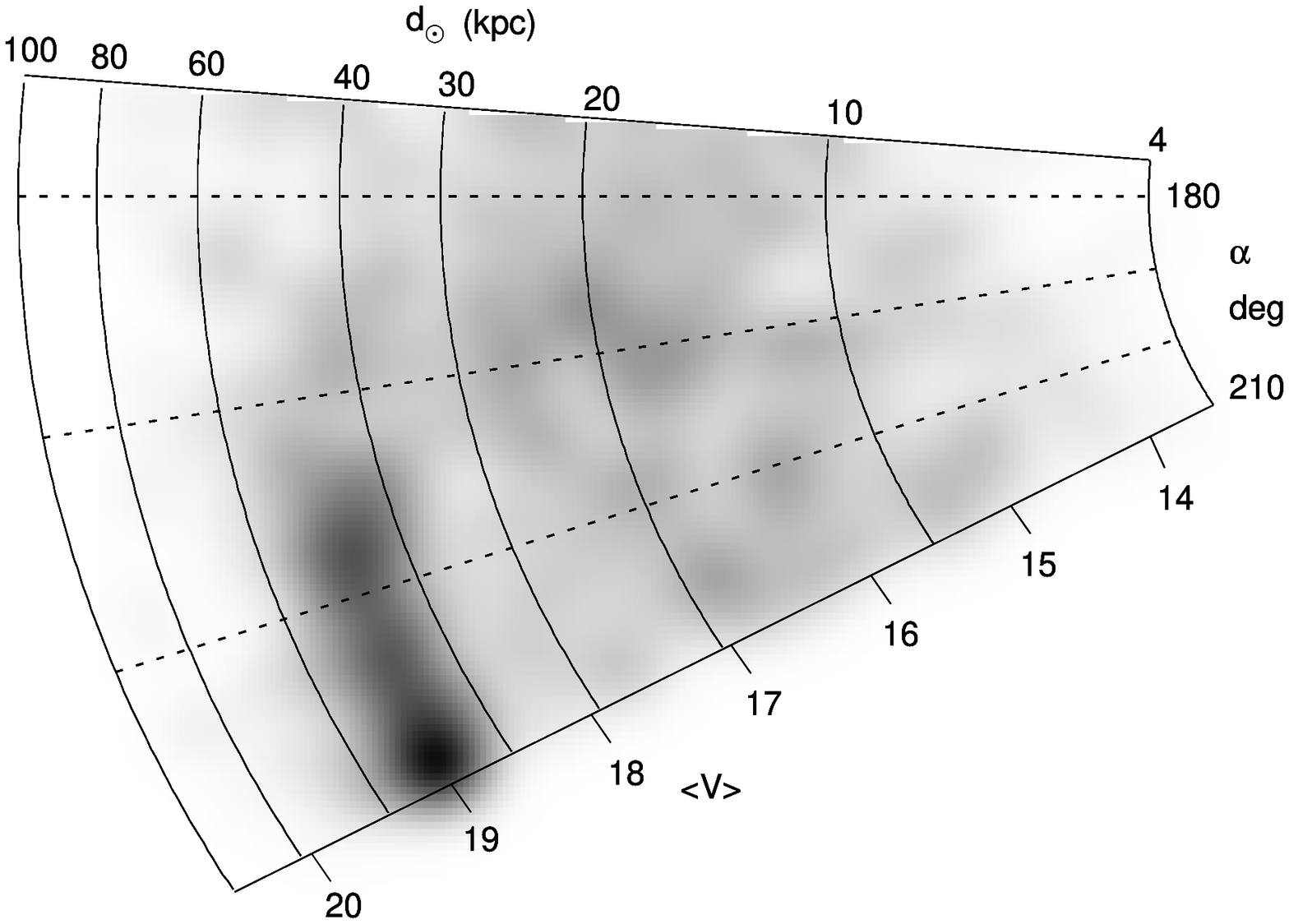}
\includegraphics[angle=270,width=0.7\textwidth]{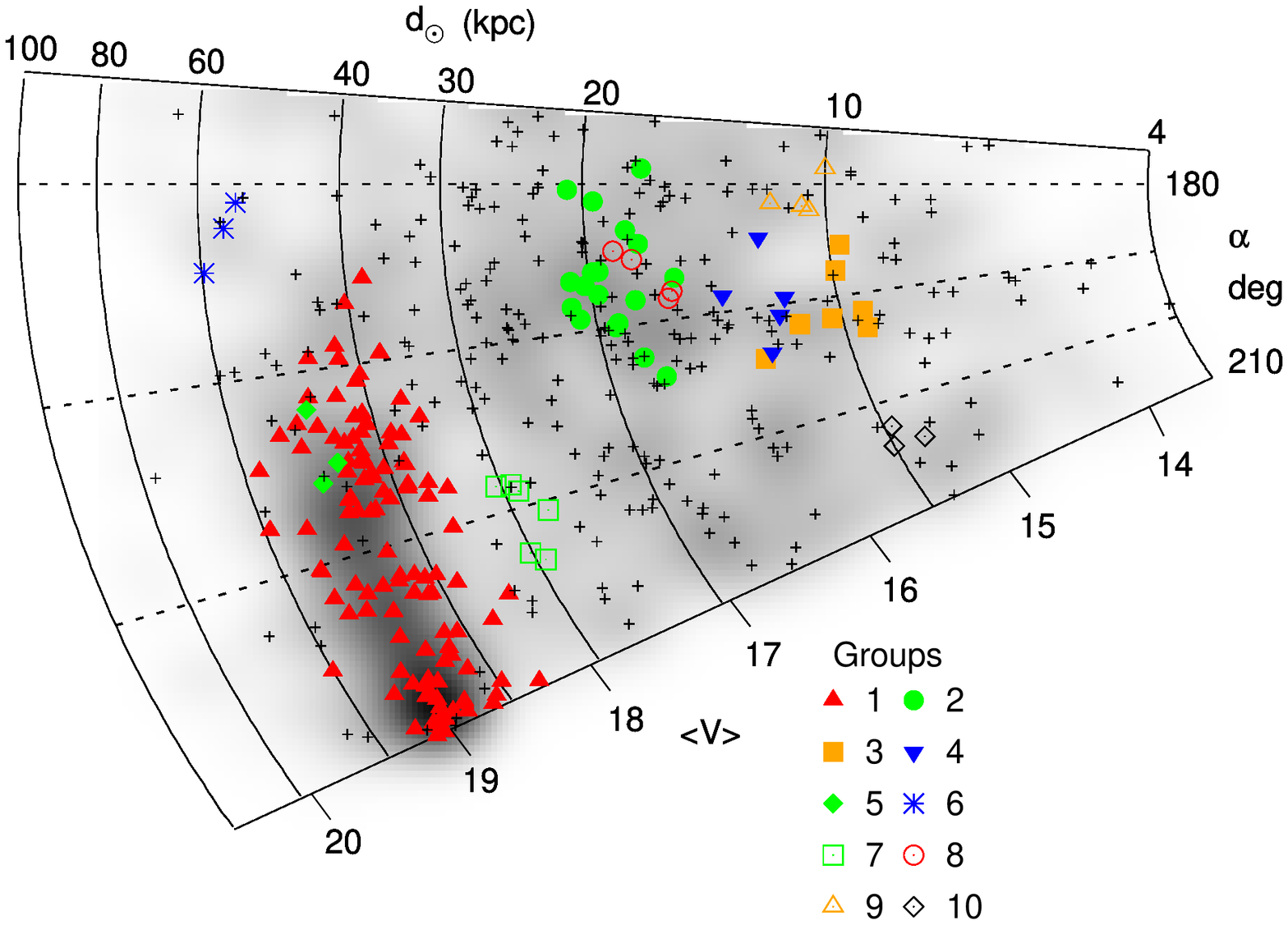}
\caption{(Top) Density distribution of the LSQ RRLS as a function of $\alpha$ and distance from the Sun, modified from \citet[][Fig 14]{zinn14}.
Regions with overdensities of RRLS are seen in darker colors. (Bottom) Same but with RR Lyrae stars with available spectrocopy overplotted as small $+$
symbols. RRLS in groups flagged as significant (Table~\ref{tab-groups}) are shown with colored symbols similar to the previous figures. 
}
\label{fig-density}
\end{figure}

\subsubsection{Relation with spatial over-densities}
In Figure~\ref{fig-density} we show the significant groups of RRLS plotted over the spatial density of RRLS in the LSQ survey 
\citep[modified from][]{zinn14}, on a scale where darker grey indicates higher density.
It is clear that the two most numerous groups of RRLS selected by our friends-of-friends algorithm agree quite well with the location of the largest densities of RRLS.  
The added information from radial velocity measurements reveals smaller groups that are not obvious in the spatial distribution.  Some of the less significant spatial overdensities (lighter levels
of grey in Figure~\ref{fig-density}) may be nothing more than random fluctuations in density because we did not find significant groups are their locations.  However, groups with small values of $|V_{gsr}|$ are not identified by our technique unless they have a large number of members or lie at large or small distances. Some of the groups in Figure~\ref{fig-allgroups} that have few members, low values of $|V_{gsr}|$ and small distances may be real, but only the addition of proper motion data and the derivation of space velocities will tell for sure.          

In agreement with the density distribution seen in the LSQ survey \citep{zinn14},  the highly significant groups (large solid symbols)
are concentrated in two ranges of distances. The ones ranging from 10 to 20 kpc agree with the upper range of distances determined for the VOD using main sequence stars \citep{juric08,bonaca12},
and we suggest that the combination of these groups are at least partially responsible for the excess of stars in this part of
the sky. The second set of groups are located between 40 and 60 kpc. Given its location and
velocity, the most numerous of these distant groups is undoubtedly the recent wrap of the leading tail of the Sgr dSph galaxy \citep[see for example][]{zinn14}. 

\begin{figure}[ht]
\plotone{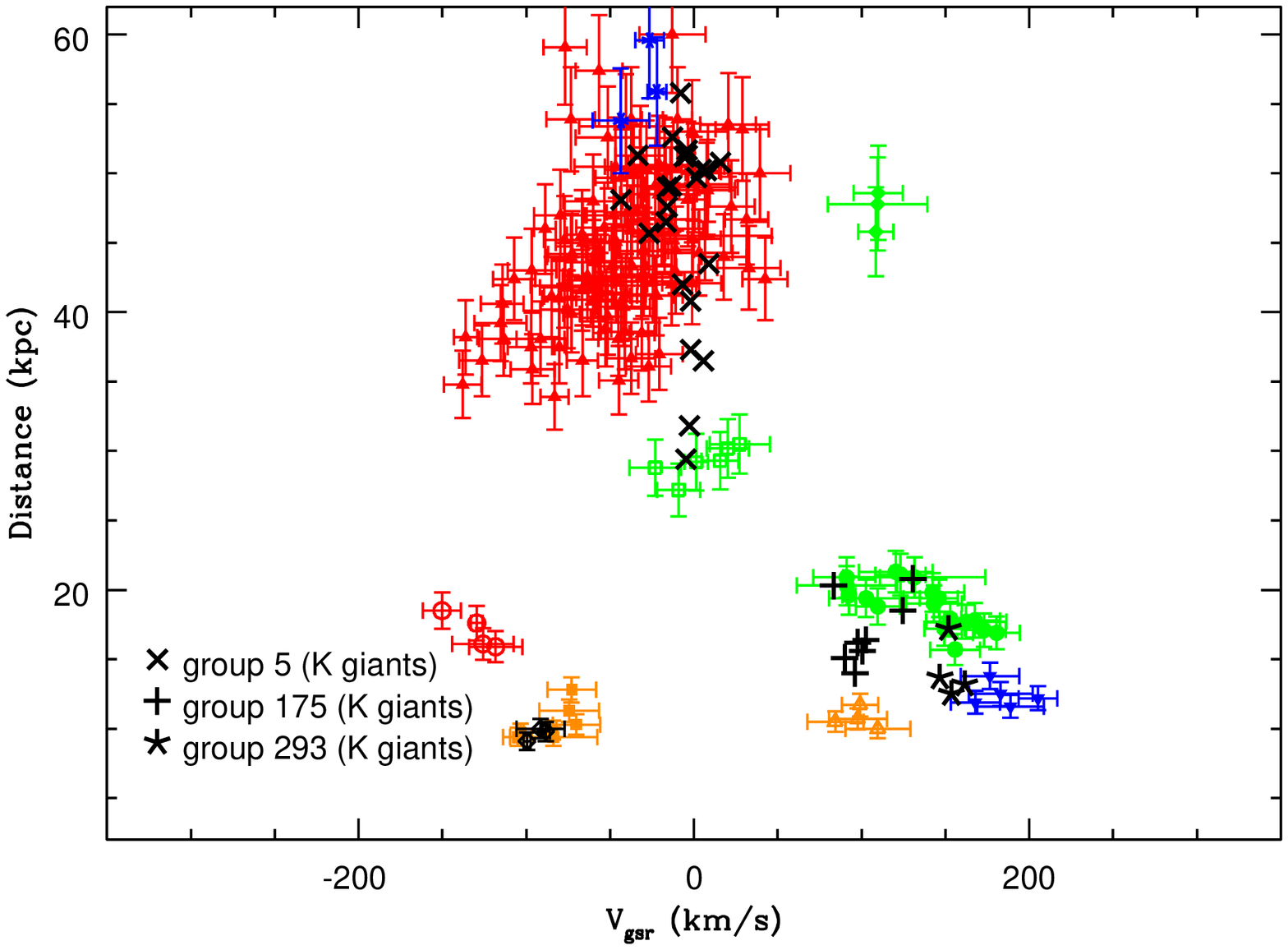}
\caption{Significant groups of RRLS in the Distance-Velocity plane are shown with colored symbols. 
Solid symbols represent the members of the significant kinematical 
groups found in the region (Table~\ref{tab-groups}) while
open symbols represent the low-significance groups.
Symbols/colors are the same as the ones shown in 
Figures~\ref{fig-allgroups} and ~\ref{fig-percentile}.
The figure shows also the location in this diagram of the three groups of K giant stars identified by \citet{janesh16} in this part of
the sky.  
}
\label{fig-Kgiants}
\end{figure}

\subsubsection{Relationship with groups of red giants}
Using a similar friends-of-friends algorithm but with a different way to calculate $4d$, 
\citet{janesh16} recently found three groups of K giants in this part of the sky and in the same range of distances.
These groups are plotted together with the significant groups of RRLS in  Figure~\ref{fig-Kgiants}.
Their group \#175 has $V_{gsr}=103$ {\kms } and a mean distance of 18 kpc (8 members), while the group
\#293\footnote{Group 293 was incorrectly labeled as 93 in Table 2 from \citet{janesh16} (W. Janesh, private communication).} 
has 4 members with mean distance of 15.1 kpc and   $V_{gsr}=153$ \kms.
There is enough overlap in location and velocity between the members of these two groups of K giants 
with two of the RRLS groups in the VOD (Groups 2 and 4) to believe they are tracing the same 
structures. 
In addition, group \#5 in \citet{janesh16} has similar properties to our Sgr group (group 1, red triangles).
The $4d$ algorithm, as implemented in \citep{janesh16}, links together stars that lie on similar lines of sight, but span a wider range of distances than our method.  This may be the 
reason why group \#5 in \citet{janesh16} spans $\sim 25$ kpc in distance and overlaps with not only our group 1 of RRLS but also with the lower significance group \# 7.

\section{Dissentangling the sub-structures in the VOD region}

\begin{figure*}[ht]
\centering
\includegraphics[angle=270,width=0.49\textwidth]{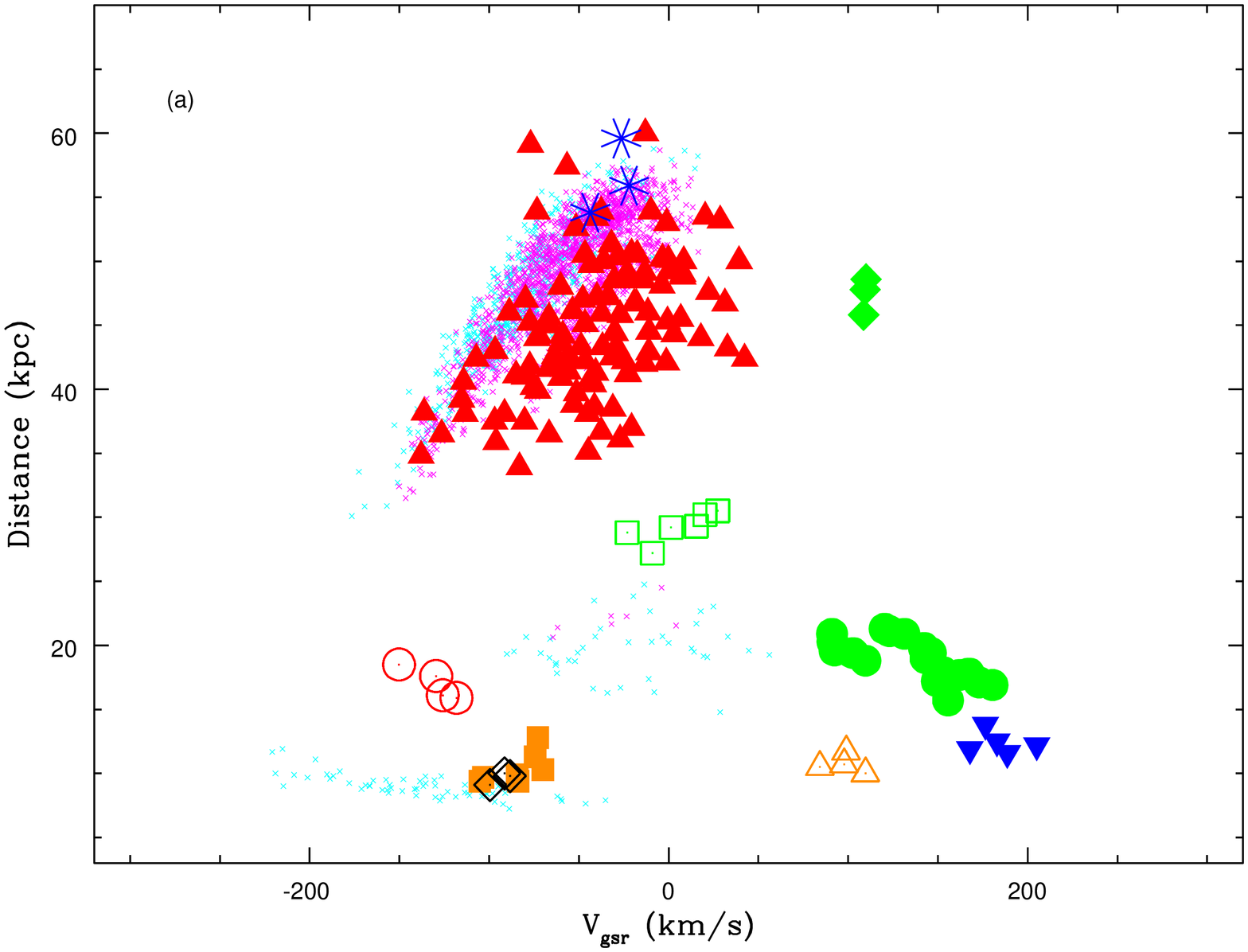}
\includegraphics[angle=270,width=0.49\textwidth]{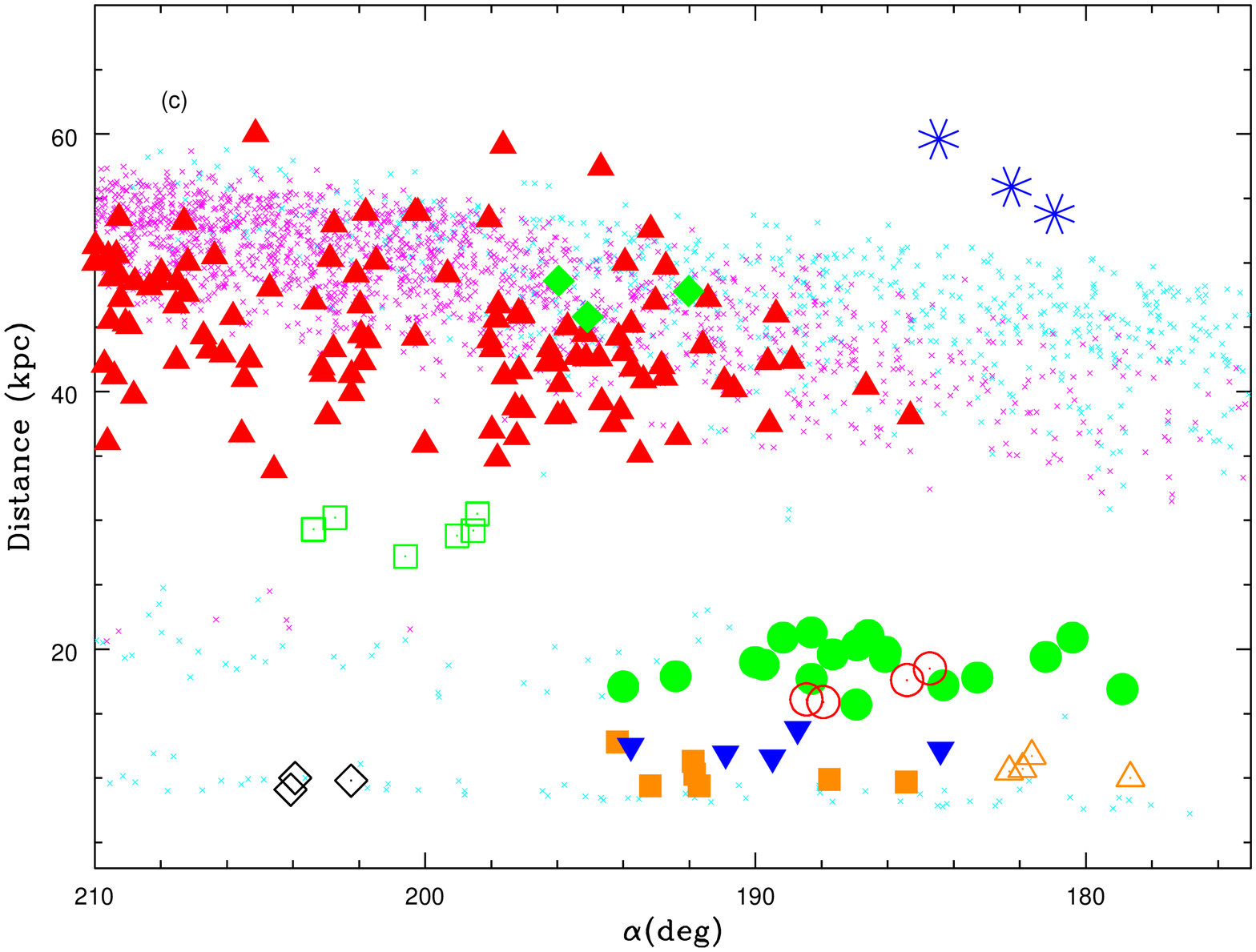}
\includegraphics[angle=270,width=0.49\textwidth]{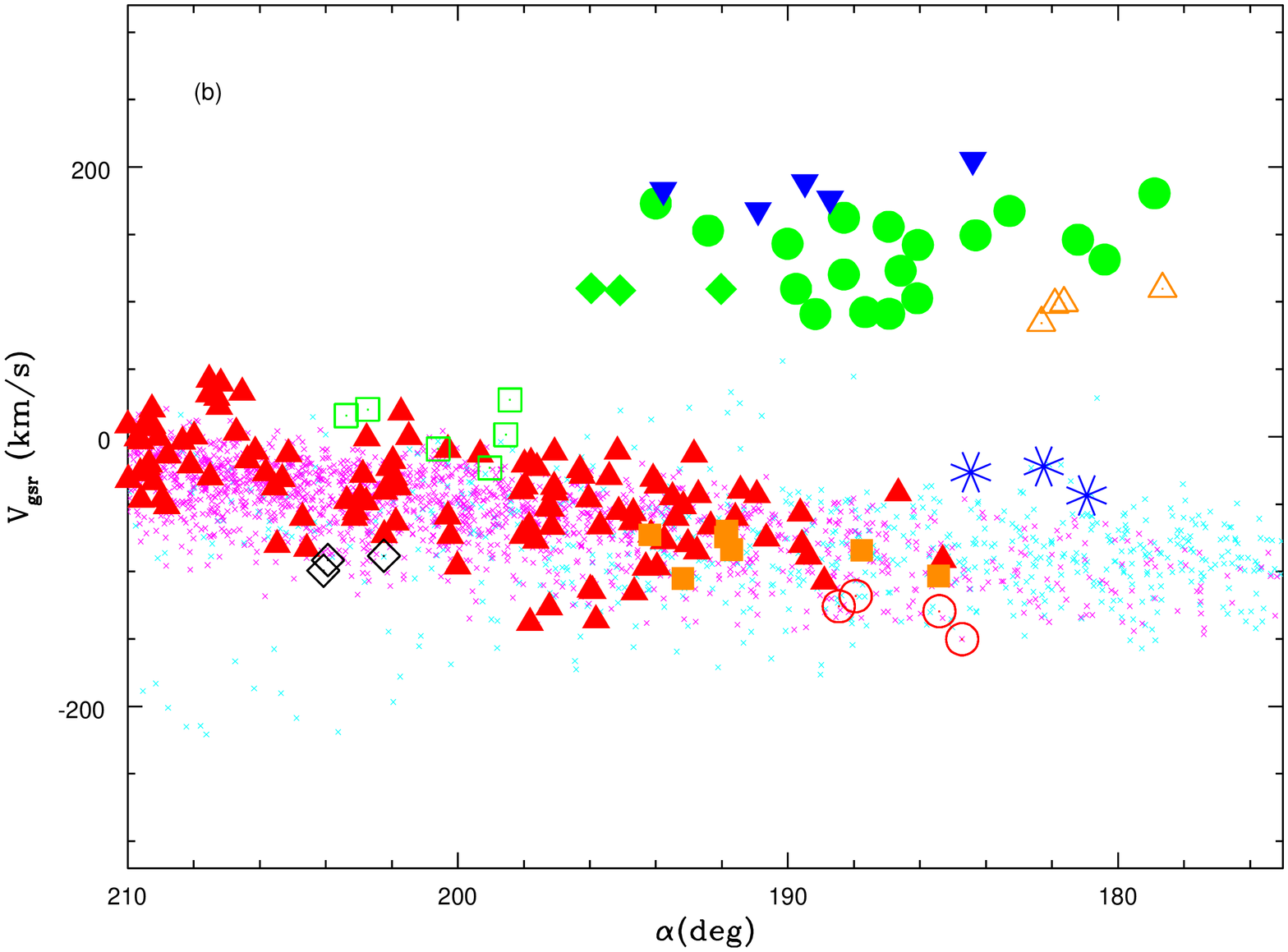}
\includegraphics[angle=270,width=0.49\textwidth]{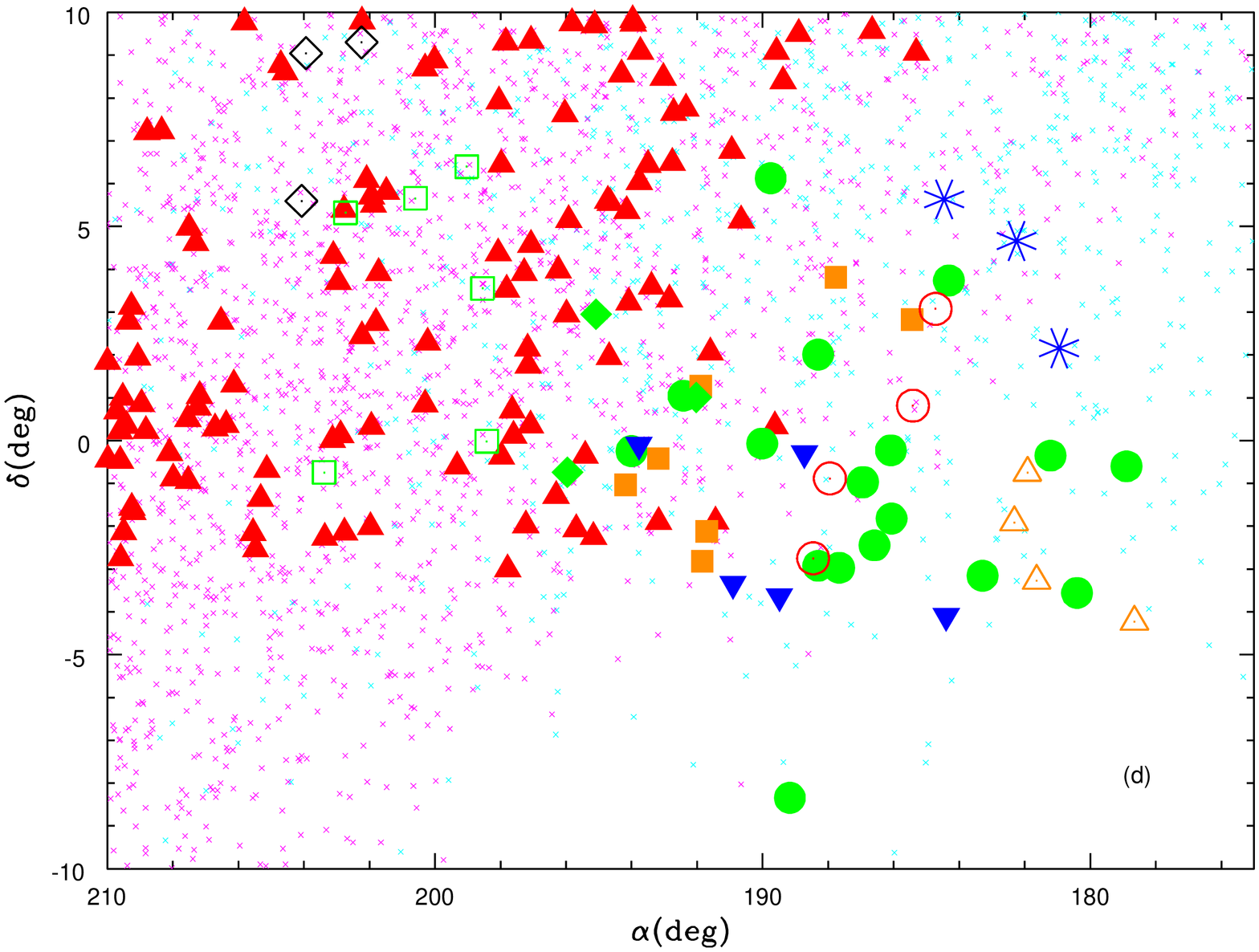}
\caption{Different views of the distribution of the RRLS studied in this work. Panel (a) Distance versus radial velocity, (b) Radial velocity versus Right Ascension, (c) Distance versus Right Ascension, and (d) Declination versus Right Ascension.
Large colored solid symbols represent the members of the significant kinematical 
groups found in the region (Table~\ref{tab-groups}).  
The large colored open symbols represent the low-significance groups.
Symbols/colors are the same as the ones shown in 
Figures~\ref{fig-allgroups} and ~\ref{fig-percentile}.
The small magenta and cyan crosses in the background represent the Sgr debris which were separated from the
main body of the galaxy 1.3-3.2 and 3.2-5.0 Gyr ago respectively, as expected from the models of \citet{law10}. 
}
\label{fig-significant_groups}
\end{figure*}

In order to understand better the extent and location of the significant groups of RRLS, we have plotted 
their location in the sky, in the Distance-Velocity diagram, and their Distance and Velocity versus right ascension in
Figure~\ref{fig-significant_groups}. These diagrams also display the expected Sgr debris in this part of the sky from the 
models of \citet{law10}. We discuss the properties of the VOD and Sgr groups below. The different views help to disentangle
the different structures, several of which lie along the same line of sight.

\subsection{the VOD}

In the distance range of the VOD, the largest group of RRLS in our study 
has 18 members (Group 2, green circles in Figure~\ref{fig-significant_groups}). This group is highly significant.
It not only has a large number of members, but its large mean 
radial velocity is also hard to reproduce in our simulations.
It also has a compact size compared to the fake groups formed in the simulations.
The mean separation of the stars from the center of the group is 2.11 kpc, and less than 10\%
of all fake groups with 18 members have mean separations equal or less than that value.
Its velocity and distance clearly corresponds with the VSS \citep{duffau06,duffau14,newberg07}.
With our new implementation of the 4d and our extended sample, this group now has twice the number of RRLS members than
\citet{duffau14}. Stars in this group strongly concentrate around 19 kpc, although there are stars between 16 and 21.3 kpc. The mean velocity is
$135.2$ \kms with a standard deviation of $28.1$ \kms. After subtracting in quadrature the mean observational error of the stars in this group ($17.5$
\kms), we obtain a true dispersion of 22 \kms. Panels b and c in Figure~\ref{fig-significant_groups} shows that there is no gradient neither in distance nor in velocity as
a function of right ascension.  There is trend of decreasing velocity with increasing distance (panel a). Among the three groups in the VOD distance range,  this is not only the most numerous group but also
it is the most extended one in 
the sky, covering from $178\degr < \alpha < 194\degr$ and $-8.4\degr < \delta < 6.1\degr$  (Panel d).  This region includes the two fields in which \citet{newberg07} observed the velocities of F-type main-sequence stars.  The most prominent peak in their velocity distribution was at $130 \pm 10$ \kms, which coincides with what we find here for Group 2, and there is also reasonable agreement in distance from the Sun as well.  The red giant group \#175 in \citet{janesh16} (see above), other red giants measured by \citet{casey12}, some main-sequence stars observed by \citet{brink10}, and a few blue horizontal branch (BHB) stars \citep{duffau14} are also likely members of the VSS.  There is little doubt that the VSS is a diffuse cloud-like halo overdensity.

Along the same line of sight of the VSS we find 4 other groups that have distances in the 
range of the VOD. 
They cover large areas of the sky, although not as large as the VSS. 
The group at 12 kpc (Group 4) is a combination of groups H and F already reported in \citet{duffau14}. No new member was found here, and indeed,
not all members in \citeauthor{duffau14} were picked here due to our stricter criteria in the separation between stars forming  a pair at short distances.  Group 4
seems to be confined to $-4\degr < \delta < 0\degr$ and it was flagged as significant because its
velocity is very high. Group 293 in \citet{janesh16} overlaps in distance and velocity with this group of RRLS but its stars are clumped around
$\delta=+10.6\degr$.

Groups 2 and 4 lie in approximately the same region of the sky, and in velocity-distance diagram, group 4 appears to be an extension of group 2 to higher $V_{gsr}$ and shorter distance.  Their separate identities could be simply a consequence of our strict criteria on group membership.  Conversely, they could truly separate and originated in separate accretion events, or they might be different wraps of the stellar streams from one event. These groups actually remain separated even if we increase the criteria to $N=2.3\sigma_v$. Measurements of the proper motions of the RRLS are required to sort this out.

Low significance group 9 lies at approximately the same distance as group 4.  Because group 9 has a significantly lower velocity, it was not linked to group 4.  While more similar to group 2 in velocity, group 9 is much in the foreground of that group.

Two other groups along the same line of sight, groups 3 and 8, have negative velocities, but quite different distances, 10.4 and 17.0 kpc, respectively.  A number of other studies of the VOD have detected negative velocity peaks \citep{newberg07, brink10, casey12, carlin12}, which overlap with the velocities of these groups of RRLS.  Given the large distance uncertainties inherent in the main-sequence stars and red giants observed in these other studies, it is not clear that these observations relate to Sgr debris at $\sim 40$ kpc, as suggested by these authors, or are a mixture Sgr debris and the same structures as RRLS groups 3 and 8. 

As shown in Figure~\ref{fig-significant_groups}, group 3 and another group of RRLS with negative velocity, group 9, overlap in distance and velocity with a weakly populated feature of the \citet{law10} model of the Sgr streams, which is due to its trailing tail.  Again, proper motion data are needed to test the reality of the RRLS groups and this possible association.

\subsection{Sgr debris}

Three high significance groups of RRLS are located at distances larger than 40 kpc. 
In the background of Figure~\ref{fig-significant_groups}, we show the location of Sgr debris in this region of the sky from the
models of \citet{law10}. Following the same color scheme in that paper, we used magenta dots to mark
particles that became gravitationally unbound in recent passages of Sgr, while cyan dots indicate particles stripped off in older
passages. 
The magenta dots are the ones that match observations from 2MASS and SDSS \citep{law10}.
Our numerous Group 1 (red triangles) closely follow the debris indicated by the magenta dots. The agreement in the sky (panel d) and
in velocity (a) is remarkably good. There seems to be a systematic offset of $\sim 5$ kpc in heliocentric distance (see panel b), with the model being 
farther away than our group. This discrepancy, which was also noticed by \citet{zinn14}, may be due to differences in the adopted distance scale between the model and our data. A similar discrepancy
was recently observed using numerous main sequence stars in the Next Generation Virgo Cluster Survey \citep[NGVS,][]{lokhorst16}.
Nonetheless, our RRLS group displays the same gradients in distance and velocity as a function of RA as the models.  In position, distance, and velocity, Group 1 agrees well with the observations by \citet{yanny09} of BHB and M giants in the Sgr stream.  The velocities of the Group 1 stars also overlap with the major peaks in the velocities of the main-sequence stars observed by \citet{brink10} and the red giant stars observed by \citet{casey12} in relatively small fields in the VOD.  Both papers identified the velocity peaks with the Sgr stream.     

Two other groups in Table 5, which have only three members each, were judged significant because they are very distant.
The most distant group (Group 6, blue asterisks) has a mean distance of 56 kpc and a velocity similar to that of the main Sgr group. It is, however, located westward of this group (see panel b and c). At its location
($\alpha\sim 182\degr$) the most recent wrap of Sgr debris, according to the model of \citet{law10}, is expected to have a very low density and be closer
than this group of RRLS by $\sim 15$ kpc without any correction for a difference in distance scale and $\sim 20$ kpc with correction.  A better match is obtained with the predicted debris from the earlier wrap (cyan dots in Figure~\ref{fig-significant_groups}), although a difference of $\sim 10$ kpc still exists. \citet{zinn14} have concluded from the spatial distribution of the LSQ RRLS that this wrap may not exist in this area of the sky.  On the basis of these factors, we conclude that Group 6 is probably not Sgr debris.   

The other distant group, Group 5 (green diamonds), overlaps in space with Group 1, which is undoubtedly Sgr debris, but its velocity is quite
different. While groups of such small numbers of stars must be viewed with some caution, if Groups 5 and 6 are indeed real, they provide additional evidence that the outer halo is laced with accretion debris, as predicted by recent models of galaxy formation \citep[e.g.][]{pillepich15,cooper15}.

A more complete discussion on the Sgr stream as seen by its RRLS in a more extended sample is 
deferred to a future publication.

\subsection{Virgo Z and related overdensities}

At $(\alpha, \delta) =  (185\fdg 077, -1\fdg 35)$, there is a candidate dwarf galaxy identified by \citet{walsh09} in the footprint of SDSS and located
at $\sim 40$ kpc from the Sun. 
The existence of this galaxy has been challenged by \citet{barbuy13} who claims that there is a distant cluster of galaxies at that position of the sky which 
may have been confused for a stellar overdensity in the shallower SDSS data. However, \citet{jerjen13} built deep CMDs in this location and in a neighbor one, at $(\alpha, \delta) =  (191\fdg 992, -0\fdg 75)$, and
concluded that both fields contain indeed a population of stars, not at 40 kpc but at 23 kpc. The fact that the main sequence observed in both fields is similar, makes 
the authors believe that they were detecting not a dwarf galaxy but and extended structure likely related to the VOD. They also ruled out the presence of a significant stellar population at 19 kpc, where
the VSS RRLS stars are concentrated. 
This non-detection of a main sequence at the distance of the VSS by \citet{jerjen13} may be due to the low surface brightness of the feature and the
small areas of the fields observed by \citet{jerjen13}.  Note that main-sequence of the VSS is well-documented by the studies of \citet{newberg02} and \citet{newberg07} in other fields.

\citet{zinn14} identified 4 RRLS in the LSQ survey which are located within 1 degree of the alleged position of Virgo Z, but at distances more compatible with the main sequence observed by \citet{jerjen13}. Those 4 stars are included in our spectroscopic sample (LSQ 515, LSQ 525, LSQ 532 and LSQ 550) and we can investigate any possible association  among them. Table~\ref{tab-VirZ} contains the distance and radial velocities obtained for these stars.

Examining Table~\ref{tab-VirZ}, it is clear that star LSQ 525 has a velocity significantly different from the others. 
However, none of the other three stars paired with each other using $4d<2\sigma v$, weakening the hypothesis that they were part of 
a stellar system in the region, and
none of these stars is a member of the significant kinematical group identified in this work (Table~\ref{tab-groups}).
Each of the three stars with negative velocities belong to separate groups among the several detected in our sample at those
distances (Figure~\ref{fig-allgroups}).
But none of those groups was flagged as significant after considering our halo simulations.
We note, however, that star LSQ 532 is a member of
the group with  $\langle V_{gsr} \rangle = -24$ \kms, $\langle d \rangle = 24$ kpc (8 members, dark green pentagons in Figure~\ref{fig-allgroups}). 
The group extends in RA between $177\degr$ and $194\degr$, and consequently, it overlaps with both of \citeauthor{jerjen13}'s detections.
This group could be the horizontal branch counterpart to \citeauthor{jerjen13}'s main sequence detection, even though it did not pass our significance criteria.  The measurement of the radial velocities of the main-sequence stars is needed to test this possibility. 

It is also possible that the main stellar population of the structures is too young to produce RRLS, and there is indication that this may be indeed the case since \citet{jerjen13} estimated an age of 8.2 Gyr and an [Fe/H]$\sim -0.7$ for the main sequence they
detected.  Few, if any, RRLS are likely to be produced by a stellar population of these attributes \citep[e.g.][]{bono16}.  Some could be still present if there is a minority population of older and more metal-poor stars.   

\begin{deluxetable}{ccc}
\tablecolumns{3}
\tablewidth{0pc}
\tablecaption{RR Lyrae variables near the location of Virgo Z \label{tab-VirZ}}
\tablehead{
ID & Distance & $V_{gsr}$ \\
    & (kpc)      & (\kms)       \\
}
\startdata
LSQ 515  & 20.2  & $-93 \pm 18$ \\
LSQ 525  & 26.0  & $238 \pm 19$ \\
LSQ 532  & 26.4  & $-23 \pm 20$ \\
LSQ 550  & 26.8  & $-65 \pm 37$ \\
\enddata
\end{deluxetable}

\section{Conclusions}

We have used 3-D maps in combination with radial velocities to identify and disentangle multiple sub-structures present in the galactic halo in the
line of sight of the VOD. Several different kinematic groups were identified in the sample of more than 400 RRLS covering the distance range from 4 to 75 kpc and more than 500 sq. degr. of the sky. The technique used to identify groups is based on the proximity of stars
in  both location in the sky and velocity, as well as on extensive simulations of halo stars that allowed us to detect cases that are 
unlikely due to random grouping. We have applied the technique to a region of the sky with a known an extensive halo feature, the VOD. However, this method can be used to explore other regions of the sky with or without known spatial overdensities. The new large-scale spectroscopic surveys such as Gaia, RAVE and APOGEE will provide interesting samples of stars to look for sub-structures in different parts of the sky. 

The two most significant features in our sample of RRLS in the VOD region are the debris from the Sgr dSph galaxy and the VSS, in agreement with previous studies \citep{duffau06, newberg07, prior09, brink10, casey12, carlin12, zinn14, duffau14, janesh16} that used RRLS, main-sequence stars, and red giants as probes.  The precise distance determinations afforded by RRLS has allowed us to identify some other potentially interesting groups of RRLS at both small and large distances, which less precise distance indicators would have blended together with the Sgr debris or the VSS.  Some of these groups lie in the distance range estimated for the VOD, which suggests that it may be a composite of several different structures, as discussed previously by several authors \citep[e.g.][]{newberg07,vivas08,duffau14}.  At large distances, there are a few, sparsely populated groups of RRLS, which if real halo substructures, suggest that the outer halo is littered with debris from accretion events.     

Although this work has expanded the area of spectroscopic study of the VOD significantly, we still cannot answer the basic questions of what is the origin of this feature and what is its
relationship (if any) with other similar cloud-like features found in the halo at similar heliocentric distances. The new generation of spectroscopic and astrometric surveys may answer these questions in the near future. 

\acknowledgements
This research was conducted as part of the Cerro Tololo Inter-American Observatory REU Program, which is supported by the National Science Foundation under grant 
AST-1062976 respectively. RZ acknowledges support from NSF grant AST-1108948.
Support
for SD is provided by the Ministry of Economy, Development, and Tourism's Millennium Science Initiative through grant IC120009, awarded to The Millennium Institute of Astrophysics, MAS.
Based in part on observations obtained at the Southern Astrophysical Research (SOAR) telescope, which is a joint project of the Ministério da Ciência, Tecnologia, e Inovação (MCTI) da República Federativa do Brasil, the U.S. National Optical Astronomy Observatory (NOAO), the University of North Carolina at Chapel Hill (UNC), and Michigan State University (MSU). 
It is also partly based on observations at Kitt Peak National Observatory, National Optical Astronomy Observatory, which is operated by the Association of Universities for Research in Astronomy (AURA) under a cooperative agreement with the National Science Foundation. 
Funding for SDSS-III has been provided by the Alfred P. Sloan Foundation, the Participating Institutions, the National Science Foundation, and the U.S. Department of Energy Office of Science. The SDSS-III web site is http://www.sdss3.org/.
SDSS-III is managed by the Astrophysical Research Consortium for the Participating Institutions of the SDSS-III Collaboration including the University of Arizona, the Brazilian Participation Group, Brookhaven National Laboratory, Carnegie Mellon University, University of Florida, the French Participation Group, the German Participation Group, Harvard University, the Instituto de Astrofisica de Canarias, the Michigan State/Notre Dame/JINA Participation Group, Johns Hopkins University, Lawrence Berkeley National Laboratory, Max Planck Institute for Astrophysics, Max Planck Institute for Extraterrestrial Physics, New Mexico State University, New York University, Ohio State University, Pennsylvania State University, University of Portsmouth, Princeton University, the Spanish Participation Group, University of Tokyo, University of Utah, Vanderbilt University, University of Virginia, University of Washington, and Yale University. We thank the anonymous referee for useful comments.

\facilities{WIYN (Hydra), SOAR}

\software{IRAF, Topcat}

\end{document}